\documentclass[aps,prd,twocolumn,superscriptaddress,nofootinbib]{revtex4-1}
\pdfoutput=1

\usepackage{amsfonts}
\usepackage{psfrag}
\usepackage{graphicx}
\usepackage{grffile}
\usepackage{cancel}
\usepackage{amsmath}
\usepackage{natbib}
\usepackage{xr}
\usepackage{hyperref}
\usepackage{verbatim}
\usepackage{epstopdf}
\usepackage{subcaption}
\usepackage{caption}
\usepackage{multirow}
\usepackage{float}
\usepackage{commath}
\usepackage{empheq}
\usepackage{booktabs,dcolumn,caption}
\hypersetup{
    bookmarks=true,         
    unicode=false,          
    pdftoolbar=true,        
    pdfmenubar=true,        
    pdffitwindow=false,     
    pdfstartview={FitH},    
    pdfauthor={Nathan Rutherford},     
    colorlinks=true,       
    linkcolor=blue,          
    citecolor=blue,        
}
\newcolumntype{d}[1]{D{.}{.}{#1}}

\RequirePackage[normalem]{ulem} 
\RequirePackage{color}\definecolor{RED}{rgb}{1,0,0}\definecolor{BLUE}{rgb}{0,0,1} 
\providecommand{\DIFaddbegin}{} 
\providecommand{\DIFaddend}{} 
\providecommand{\DIFdelbegin}{} 
\providecommand{\DIFdelend}{} 
\providecommand{\DIFaddbeginFL}{} 
\providecommand{\DIFaddendFL}{} 
\providecommand{\DIFdelbeginFL}{} 
\providecommand{\DIFdelendFL}{} 
\newcommand{\DIFscaledelfig}{0.5}
\RequirePackage{settobox} 
\RequirePackage{letltxmacro} 
\newsavebox{\DIFdelgraphicsbox} 
\newlength{\DIFdelgraphicswidth} 
\newlength{\DIFdelgraphicsheight} 
\LetLtxMacro{\DIFOincludegraphics}{\includegraphics} 
\newcommand{\DIFaddincludegraphics}[2][]{{\color{blue}\fbox{\DIFOincludegraphics[#1]{#2}}}} 
\newcommand{\DIFdelincludegraphics}[2][]{
\sbox{\DIFdelgraphicsbox}{\DIFOincludegraphics[#1]{#2}}
\settoboxwidth{\DIFdelgraphicswidth}{\DIFdelgraphicsbox} 
\settoboxtotalheight{\DIFdelgraphicsheight}{\DIFdelgraphicsbox} 
\scalebox{\DIFscaledelfig}{
\parbox[b]{\DIFdelgraphicswidth}{\usebox{\DIFdelgraphicsbox}\\[-\baselineskip] \rule{\DIFdelgraphicswidth}{0em}}\llap{\resizebox{\DIFdelgraphicswidth}{\DIFdelgraphicsheight}{
\setlength{\unitlength}{\DIFdelgraphicswidth}
\begin{picture}(1,1)
\thicklines\linethickness{2pt} 
{\color[rgb]{1,0,0}\put(0,0){\framebox(1,1){}}}
{\color[rgb]{1,0,0}\put(0,0){\line( 1,1){1}}}
{\color[rgb]{1,0,0}\put(0,1){\line(1,-1){1}}}
\end{picture}
}\hspace*{3pt}}} 
} 
\LetLtxMacro{\DIFOaddbegin}{\DIFaddbegin} 
\LetLtxMacro{\DIFOaddend}{\DIFaddend} 
\LetLtxMacro{\DIFOdelbegin}{\DIFdelbegin} 
\LetLtxMacro{\DIFOdelend}{\DIFdelend} 
\DeclareRobustCommand{\DIFaddbegin}{\DIFOaddbegin \let\includegraphics\DIFaddincludegraphics} 
\DeclareRobustCommand{\DIFaddend}{\DIFOaddend \let\includegraphics\DIFOincludegraphics} 
\DeclareRobustCommand{\DIFdelbegin}{\DIFOdelbegin \let\includegraphics\DIFdelincludegraphics} 
\DeclareRobustCommand{\DIFdelend}{\DIFOaddend \let\includegraphics\DIFOincludegraphics} 
\LetLtxMacro{\DIFOaddbeginFL}{\DIFaddbeginFL} 
\LetLtxMacro{\DIFOaddendFL}{\DIFaddendFL} 
\LetLtxMacro{\DIFOdelbeginFL}{\DIFdelbeginFL} 
\LetLtxMacro{\DIFOdelendFL}{\DIFdelendFL} 
\DeclareRobustCommand{\DIFaddbeginFL}{\DIFOaddbeginFL \let\includegraphics\DIFaddincludegraphics} 
\DeclareRobustCommand{\DIFaddendFL}{\DIFOaddendFL \let\includegraphics\DIFOincludegraphics} 
\DeclareRobustCommand{\DIFdelbeginFL}{\DIFOdelbeginFL \let\includegraphics\DIFdelincludegraphics} 
\DeclareRobustCommand{\DIFdelendFL}{\DIFOaddendFL \let\includegraphics\DIFOincludegraphics} 

\newcommand{\Msun}{\mathrm{M}_\odot}
\newcommand{\chistar}{\chi^*}

\newcommand{\phimu}{\phi_\mu}

\newcommand{\phinaught}{\phi_0}
\newcommand{\mchi}{m_\chi}
\newcommand{\mphi}{m_\phi}
\newcommand{\w}{\omega}

\begin{document}

\title{Constraining bosonic asymmetric dark matter\\ with neutron star mass-radius measurements}
\author{Nathan Rutherford}
\email[Corresponding author: ]{nathan.rutherford@unh.edu}
\affiliation{Department of Physics and Astronomy, University of New Hampshire, Durham, New Hampshire 03824, USA}
\author{Geert Raaijmakers}
\email{g.raaijmakers@uva.nl}
\affiliation{GRAPPA, Anton Pannekoek Institute for Astronomy and Institute of High-Energy
Physics, University of Amsterdam, Science Park 904, 1098 XH Amsterdam, The
Netherlands}
\author{Chanda Prescod-Weinstein}
\email{chanda.prescod-weinstein@unh.edu}
\affiliation{Department of Physics and Astronomy, University of New Hampshire, Durham, New Hampshire 03824}
\author{Anna Watts}
\email{A.L.Watts@uva.nl}
\affiliation{Anton Pannekoek Institute for Astronomy, University of Amsterdam,
Science Park 904, 1090GE Amsterdam, The Netherlands}
\bibliographystyle{unsrtnat}
\begin{abstract}
Neutron stars can accumulate asymmetric dark matter (ADM) in their interiors, which affects the neutron star's measurable properties and makes compact objects prime targets to search for ADM. In this work, we use Bayesian inference to explore potential neutron star mass-radius measurements, from current and future x-ray telescopes, to constrain the bosonic ADM parameters for the case where bosonic ADM has accumulated in the neutron star interior. We find that the current uncertainties in the baryonic equation of state do not allow for constraints on the ADM parameter space to be made. However, we also find that ADM cannot be excluded and the inclusion of bosonic ADM in neutron star cores relaxes the constraints on the baryonic equation of state space. If the baryonic equation of state were more tightly constrained independent of ADM, we find that statements about the ADM parameter space could be made. In particular, we find that the high bosonic ADM particle mass ($m_\chi$) and low effective self-interaction strength ($g_\chi/m_\phi)$ regime is disfavored due to the observationally and theoretically motivated constraint that neutron stars must have at least a mass of $1 \, \Msun$. However, within the remaining parameter space, $m_\chi$ and $g_\chi/m_\phi$ are individually unconstrained. On the other hand, the ADM mass-fraction, i.e., the fraction of ADM mass inside the neutron star, can be constrained by such neutron star measurements.
\end{abstract}

\maketitle
\section{\label{intro}Introduction}
Neutron stars have been of great interest to astronomers because of the rich phenomena they produce, allowing us to probe, for example, strong gravity, cosmology, and heavy element enrichment. For instance, neutron stars can be a source of continuous gravitational waves, which can provide insight into their interiors. Signals from their mergers can also be used in determining the Hubble constant \cite{Sieniawska2019,Chen2018}. Neutron stars have additionally been of great interest to nuclear physicists because the microphysical behavior of ultradense neutron-rich matter is poorly understood. Neutron stars may contain exotic states of matter, such as hyperons or deconfined quarks \cite{Oertel_2017,Burgio_2021,Han2022}. The effects of the hypothetical components that comprise neutron star interiors are parametrized by the equation of state (EoS), which has traditionally been understood to only reflect baryonic components. The EoS both theoretically captures information about neutron star interiors and can be deduced from measurable properties of neutron stars. Thus, understanding the potential impact of any given component is both theoretically and observationally important.

Recent work has shown dark matter, too, can have a significant impact on currently measurable properties of neutron stars, namely the masses, radii, and tidal deformabilities, with implications for the EoS. Dark matter cores in neutron stars have been shown to reduce the radii, masses, and tidal deformabilities of the stars \cite{Ellis2018,Kain2021,Sagaun_2020,Sagun_2022_Boson,Sagun_2021_fermion}. If dark matter forms a halo around neutron stars, the tidal deformabilities and the total masses of the stars have been shown to increase \cite{Nelson_2018,Kain2021,Sagaun_2020,Sagun_2021_fermion,Sagun_2022_Boson}. Thus, if dark matter is present in neutron stars, it must be accounted for in estimates of the measurable properties of neutron stars \cite{Lavallaz2010,Nelson_2018,Kain2021,Ellis2018,Sen2021,Guha2021,Sagaun_2020,Husain2021,Sagun_2021_fermion,Sagun_2022_Boson,Miao_2022,Das2020,Das_2021,Das_2021pt2,Das_2022,Apran_2019,Collier2022,Leung2022}. Dark matter can also accumulate enough mass inside a neutron star to form a black hole, which can tightly constrain dark matter models through observations of old neutron stars \cite{Khlopov1985,Bertone2008,Kouvaris2011,Kouvaris2011_pt2,McDermott2012}. When specific assumptions are made about the EOS, the dark matter particle mass, self-interaction strength, and mass-fraction can also potentially be constrained using neutron star and gravitational wave (GW) measurements \cite{Apran_2019,Apran2020,Das_2021,Sen2021,Guha2021,Miao_2022,Sagun_2022_Boson}. Specific calculations include placing mass-fraction constraints on sub-GeV bosonic dark matter particles \cite{Sagun_2022_Boson}, demonstrating that a stiffer baryonic matter EoS with dark matter can evade constraints that the baryonic matter EoS alone cannot achieve \cite{Apran_2019,Das_2021,Sen2021,Guha2021}, and calculating Bayesian parameter estimations of the dark matter parameter space \cite{Apran2020,Miao_2022}.

To quantify the effects of dark matter on neutron stars, we calculate the percent change in mass and radius between a baryonic neutron star with and without a dark matter core using our algorithm described in Sec.~\ref{eosnumerical}. For example, if a $2.3 \, \Msun$ purely baryonic neutron star with radius of 10.7 km is compared to a baryonic neutron star with an identical central baryonic energy density and a dark matter core of $7\%$ mass-fraction, the radius of the dark matter admixed neutron star is reduced at the $6 \%$ level while the gravitational mass is decreased at the $9 \%$ level. The dark matter core increases the compactness of the neutron star within the dark matter core radius, which reduces the neutron star's gravitational mass \cite{Ellis2018,Sagaun_2020,Sagun_2021_fermion,Sagun_2022_Boson,Apran_2019}. This demonstrates that the presence of dark matter can have an observable impact on neutron stars, suggesting a need to account for this possibility in data analyses. 

Searching for observable effects of the impact of dark matter on the neutron star interior requires measurements of the mass and radius of the neutron star. These measurements can be used to characterize the mass-radius relation, which is the relation of all possible, stable neutron star masses and radii that can correspond to a specific EoS. Calculating the mass-radius relation makes it possible to obtain hypothetical EoSs of the neutron star \cite{Lindblom_1992}. Mass and radius measurement techniques have been advanced by NICER (the Neutron Star Interior Composition Explorer), an x-ray telescope on the International Space Station \cite{Gendreau2017}. NICER uses Pulse Profile Modeling \citep[PPM, see][and references therein]{Watts2019}, a technique that exploits relativistic effects on x-rays emitted from the hot magnetic polar caps of millisecond pulsars, i.e., neutron stars that emit beams of radiation from their magnetic poles and have a rotational period less than ten milliseconds. A pulse profile is a rotationally phase-resolved x-ray count spectrum. PPM involves a Bayesian inference of this data for a generative relativistic ray-tracing model of the thermal emission from hotspots on a neutron star's surface. This analysis delivers the posterior probability distributions of the various model parameters, such as a neutron star's mass and radius and a map of the hot emitting regions, which form as magnetospheric particles slam into the stellar surface \cite{Watts2019}. To date, NICER has delivered PPM mass and radius inferences for two neutron stars: PSR J0030+0451 \citep{Riley2019,Miller2019} and PSR J0740+6620 \citep{Riley2021,Miller2021,Salmi2022}. These mass-radius measurements have been incorporated into numerous EoS studies, both from the NICER team \citep{Raaijmakers2019,Raaijmakers2020,Raaijmakers2021,Miller2019,Miller2021} and other groups \citep[see for example][]{JieLiJ21,Legred21,Pang21,TangSP21,Annala2022,Biswas2022}. 

NICER anticipates delivering improved mass-radius constraints for the sources already analyzed, in addition to results for at least three more neutron stars, in the near future. In the longer term, plans are being developed for large area x-ray spectral-timing missions that will be able to carry out PPM for a larger population of neutron stars. These include the NASA probe-class mission concept STROBE-X (the Spectroscopic Time-Resolving Observatory for Broadband Energy x-rays, \cite{Ray2019}) and the Chinese-European mission concept eXTP (the enhanced x-ray Timing and Polarimetry mission, \cite{extp}).  

Although NICER, STROBE-X, and eXTP will perform PPM on compact objects, current PPM techniques do not account for a possible dark matter component in neutron stars. Dark matter admixed neutron stars exhibit two possible spatial regimes: a dark matter core inside the neutron star, and a dark matter halo extending through and beyond the baryonic surface of the neutron star. A dark matter core is defined when the dark matter radius, $R_\chi$, is less than the baryonic radius, $R_B$. Dark matter halos are defined when the dark matter radius is greater than the baryonic radius, i.e, $R_\chi > R_B$. Capturing the impact of dark matter on neutron star structure alters how PPM is executed, and will certainly necessitate modification of the ray-tracing models currently being used by the NICER collaboration. For example, the existence of any halo will modify the exterior space-time. In addition, dark matter could modify the universal relations that are relied on to model the oblateness of the star (another factor affecting the space-time) \cite{Morsink2007,AlGendy2014,Miao_2022,DiGiovanni_2022}.

Two studies, \cite{Apran2020,Miao_2022}, have used PPM measurements from NICER, or GW measurements from LIGO/VIRGO, in the context of dark matter admixed neutron stars. In \cite{Apran2020}, the authors performed a Bayesian analysis on a fermionic dark matter model using a fixed baryonic matter EoS. Their inference is based on the posterior mass measurement of PSR J0348+0432 from \cite{Antoniadis_2013} and the tidal deformability constraints from GW170817 in \cite{GW170817eos}. The Bayesian analysis of \cite{Apran2020} showed the fermionic dark matter EoS parameters are independent of each other and well constrained. In \cite{Miao_2022}, the Bayesian analysis takes into account the fermionic model described in \cite{Nelson_2018} with a fixed baryonic matter EoS and is restricted to dark matter cores and diffuse dark matter halos. The data points used for the analysis were the mass and radius measurements of two NICER pulsars PSR J0740+6620, and PSR J0030+0451. For the halos, \cite{Miao_2022} found that the fermionic particle mass is constrained to be less than $1.5$ GeV. For cores, they found that the observational data favors a particle mass around $0.6$ GeV. However, \cite{Miao_2022} did not determine constraints on the ADM self-interaction strength and mass-fraction. To constrain dark matter in and around neutron stars, both studies used Bayesian parameter estimation with PPM or GW measurements and a fixed baryonic EoS.

Here we take a different approach since the interior of neutron stars is not well understood. The most conservative approach to a Bayesian analysis of neutron stars is by allowing all parameters in the EoS model to vary. In this work, we use the mass-radius measurements of simulated neutron stars to infer the optimal combination of baryonic matter EoS and admixed dark matter in neutron star cores. We assume the \textit{bosonic} asymmetric dark matter (ADM) model described in \cite{Nelson_2018}. One major objective is to demonstrate the possibility of constraining the properties of bosonic ADM, e.g., the possible particle mass, mass-fraction of the total accumulated ADM mass inside the neutron stars, and the effective interaction strength of ADM. Knowing the possible constraints on bosonic ADM will allow us to better understand the composition of neutron stars. The other objective is to characterize the effects on the derived uncertainties of the baryonic matter EoS when including ADM inside neutron stars. We study two possible scenarios of future mass-radius measurements, the first being the \textit{Future} scenario and the other the \textit{Future-X} scenario. The \textit{Future} case is modeled after a potential end of mission scenario for NICER and takes into account six simulated mass-radius measurements. The \textit{Future-X} scenario is modeled using six possible STROBE-X sources. These are modeled at lower uncertainties than the \textit{Future} case since the STROBE-X mission is expected to provide tighter measurements than NICER \cite{Ray2019,STROBEX2}. This work shows that the baryonic EoS uncertainties are relaxed if the possibility of ADM cores in neutron stars is accounted for. Additionally, this work finds that, if the baryonic EoS is more tightly constrained than it presently is, the possible future NICER and STROBE-X mass-radius measurements can constrain the ADM mass-fraction and the ratio of the bosonic ADM particle mass and effective self-repulsion strength. We do not find that NICER or STROBE-X can constrain the bosonic ADM particle mass or effective self-repulsion strength using mass-radius measurements. 

This paper is organized as follows. In Sec.~\ref{ADMBackground}, we discuss the background on ADM. In Sec.~\ref{twofluidTOV}, we motivate the two-fluid TOV equations and define the baryonic matter and ADM EoSs. In Sec.~\ref{eosnumerical}, we introduce the baryonic matter and ADM equation of states and their numerical implementation in the two-fluid TOV equations. In Sec.~\ref{Best Case NICER and STROBE-X Scenarios}, we describe our inference methods, the constraints on the ADM EoS parameter space, selection of simulated sources, and how we conducted the Bayesian parameter estimation for both the \textit{Future} and \textit{Future-X} scenarios. Lastly, in Sec.~\ref{Conclusion}, we discuss our results. Throughout this work, we use the metric signature $diag(-,+,+,+)$.

\section{Background: Asymmetric Dark Matter\\}\label{ADMBackground}
 Broadly, the ADM model is motivated by the observation that the mass density of dark matter in the Universe is only approximately five times greater than that of baryonic matter \cite{Petraki2013}. The similarity in the observed density of dark matter to that of baryonic matter suggests a strong connection in the cosmic history between them. This connection suggests that, like the baryon asymmetry, at some time in the early universe there was a tiny excess of dark matter particles over antidark matter particles and the dark matter particles today constitute the excess after all of the antidark matter particles were annihilated \cite{Petraki2013,Zurek2013}. This ``dark asymmetry'' would allow for both significant repulsive self-interactions and small attractive interactions with baryonic matter.

The \cite{Nelson_2018} ADM model is a MeV-GeV mass-scale bosonic dark matter particle with a repulsive self-interaction. Two identical particles self-interact when they scatter off one another through the exchange of a gauge boson, i.e., a force carrier. This exchange causes the particles to attract or repel each other. In the \cite{Nelson_2018} ADM model, the repulsive self-interaction arises from the exchange of an eV-MeV mass-scale vector gauge boson which also carries the Standard Model baryon number. The vector gauge boson carries the Standard Model baryon number to create the ``dark asymmetry'' mentioned previously.

The action of \cite{Nelson_2018}, in units of $\hbar = c = 1$, is expressed as
\begin{multline}\label{action}
    S = -\int d^4x \sqrt{-g} \Big( D_\mu^* \chistar D^\mu \chi  + m_\chi^2 \chistar\chi +\frac{1}{2} m_\phi^2 \phi_\mu \phi^\mu\\
    + \frac{1}{4}Z_{\mu\nu} Z^{\mu \nu} - g_B \phi_\mu J_B^\mu \Big),
\end{multline}
where $g $ is the determinant of the metric, $\chi$ is the charged bosonic ADM field, $\chistar$ is the anti-ADM field, $m_\chi$ is the mass of the bosonic ADM field, $\phi_\mu$ is the vector boson field of the ADM mediator, $m_\phi$ is the mass of the vector boson field, $Z_{\mu \nu} = \nabla_{[\mu}\phi_{\nu]} = \nabla_\mu \phi_\nu - \nabla_\nu \phi_\mu$ is the field strength tensor of $\phi_\mu$, $\nabla_\mu$ is the covariant derivative, $g_B$ is the interaction strength of $\phi_\mu$ with the Standard Model baryon number current\footnote{See \cite{Bass2004} for a detailed discussion of the Standard Model baryon number current.} $J^\mu_B$, $D_\mu  = \nabla_\mu + i g_\chi \phi_\mu$, and $g_\chi$ is the interaction strength of $\chi$ with the $\phi_\mu$ vector field. This model considers only the necessary interactions of the bosonic ADM model: the repulsive self-interactions of ADM, the minimal interaction of ADM to gravity, and the interaction of ADM with baryons.

We make a few approximations that simplify deriving the equations of motion. \cite{Reddy2016} shows that $g_B$ is constrained from measurements of supernova SN1987A to be 
\begin{equation}
    g_B \leq 10^{-10}. 
\end{equation} 
This constraint was determined by requiring the energy radiated due to the possible production of $\phi_\mu$, from nucleon-nucleon Bremsstrahlung reactions in proto-neutron stars, be consistent with the neutrino data of SN1987A. Although robust calculations of this inequality have yet to be done, we follow \cite{Nelson_2018} in assuming $g_B \ll g_\chi$. Making this assumption allows us to ignore the interactions between ADM and baryonic matter, and we can neglect the $g_B \phimu J_B^\mu$ term in the action \cite{Nelson_2018}. We also assume the spacetime is flat because the effects of gravity are negligible relative to the inverse length scales of neutron stars \cite{Sagaun_2020}. Assuming the spacetime to be flat implies that in our chosen coordinate system the determinant of the metric, $g$, is $-r^4sin^2(\theta)$.

With these approximations, the equations of motion are 
\begin{align}
    &\left[D_\mu D^\mu - m_\chi^2 \right] \chi = 0 \label{3}\\
    &\left[D_\mu^* D^{*\mu} - m_\chi^2 \right] \chistar = 0 \label{4}\\
    &\nabla_\mu Z^{\mu \nu} + i g_\chi \left[\chistar D^\nu \chi-(D^{*\nu} \chistar)\chi\right]-m_\phi^2 \phi^\nu = 0 \label{5}.
\end{align}
From the equations of motion one can employ the mean-field approximation to arrive at the ADM EoS which will be described in more detail in Sec.~\ref{eosnumerical}.

The energy of $\chi$ and $\chistar$ in a hot dense background, such as the background from proto-neutron stars, can also be obtained using the equations of motion. The conditions of the background from a proto-neutron star can produce neutron Bremsstrahlung reactions of ADM \cite{Nelson_2018,Ellis2018}. The neutron Bremsstrahlung converts the kinetic energy of two neutrons scattering to the gauge boson $\phi_\mu$, i.e., $NN \longrightarrow NN \phi_\mu$, where $NN$ is the pair scattering neutrons. Since $\phi_\mu$ interacts strongly with ADM and weakly with the Standard Model baryon number current, the Bremsstrahlung process causes the rate of the reaction $NN \rightarrow NN \chi \chistar$ to proceed at a comparable pace. In order to expel the anti-ADM, one must assume that ADM is attracted to baryons and anti-ADM is repulsed. It is due to the asymmetry in the energies of ADM and anti-ADM, i.e., $\Delta E = E_{\Bar{\chi}} - E_\chi$, that expels the anti-ADM and leaves the ADM inside the proto-neutron star. Neutron Bremsstrahlung is enhanced for young neutron stars because they provide a sufficiently hot and dense background. Newly born neutron stars have a temperature of approximately $5.8 \times 10^{11} \, K$ which can lead to $\approx 0.02 M_{NS}$ ADM accumulation in the interior, where $M_{NS}$ is the mass of the neutron star \cite{Ellis2018}. This mechanism would give all neutron stars comparable ADM mass-fractions with small differences related to the neutron star mass because this reaction is dependent upon the constant microphysical properties of neutron matter and ADM \cite{Nelson_2018}. 

Another mechanism whereby ADM can accumulate in neutron stars is the conversion of the neutron to scalar ADM. The conversion of neutrons to scalar ADM, $N \longrightarrow \chi$, can occur because neutrons in compact objects can reach Fermi momenta large enough to allow for the conversion reaction \cite{Ellis2018}. For neutron stars with an age $\sim 10$ Gyr, this process allows for ADM to reach masses of $\approx 0.05 M_{NS}$ \cite{Ellis2018}. Similar to the neutron Bremsstrahlung of ADM mechanism, neutron conversion to scalar ADM would also give all neutron stars with equivalent mass-fractions
\cite{Ellis2018,Nelson_2018}. Other possible scenarios, such as the collapse of a star to a neutron star from an ADM minihalo that has been accumulated from the local dark matter density and supernovae of supermassive stars which can form an ADM core inside neutron stars, have been explored in \cite{Deliyergieyev_2019} and \cite{Kouvaris_2010}, respectively.  The previous two accumulation methods of \cite{Deliyergieyev_2019, Kouvaris_2010} would result in a variability in the ADM mass-fraction due to their dependence on the local dark matter density. However, if ADM is accreted,  it has been shown that the upper bound on the accreted dark matter mass is between $10^{-5}-10^{-8} \mathrm{\Msun}$ \citep[ see][and references therein]{Sagun_2022_Boson}. To achieve higher mass-fractions of dark matter than what is possible with neutron Bremmstrahlung of ADM and the conversion of neutrons to scalar ADM, accumulation methods, such as the absorption of dark matter stars by baryonic matter or the accretion of baryonic matter on to a dark matter core, have been explored in \cite{Ellis2018} and \cite{Sagun_2022_Boson}.
\section{\label{twofluidTOV}Two-fluid TOV equations}
Typically, neutron star interiors are modeled using a hypothetical baryonic matter equation of state and solving the Tolman-Oppenheimer-Volkoff (TOV) equations for the pressure, mass, and radius given a central density \cite{Oppenheimer1939,Tolman1939}. The mass-radius relation can be calculated from the TOV equations by varying the central density \cite{Lindblom_1992}. The structure of a neutron star with an ADM component is modeled using the two-fluid TOV equations \cite{Sandin2009}, which allows for baryonic matter and dark matter to be treated as two separate fluids, and can be solved for the mass, radius, and pressure profiles for both fluids using an appropriate equation of state. The two-fluid TOV is an appropriate treatment for studying ADM in neutron stars because the dominant interfluid interaction between baryonic matter and ADM is gravitational. In the two-fluid TOV equations, the pressures and masses of both the baryonic matter and ADM fluids are simultaneously solved numerically until either one of the fluid pressures goes to zero. Then the integration is broken and restarted using the single-fluid TOV equations for the remaining fluid. Here we follow the derivation in \cite{Carroll2004} and derive the TOV equations from Einstein's field equations.

First, consider a general static, spherically symmetric metric 
\begin{equation}\label{metric}
    ds^2 = -c^2 e^{2\alpha(r)}dt^2 + \left[\frac{c^2r}{rc^2-2GM(r)} \right] dr^2 +r^2 d\Omega^2.
\end{equation}
If we model the neutron star as a perfect fluid, then the energy-momentum tensor takes the form
\begin{multline}
    T_{\mu\nu}=
    diag\Big(e^{2\alpha(r)}\frac{\epsilon}{c^2}, \left[\frac{c^2r}{rc^2-2GM(r)} \right] p, \\ r^2 p , \, r^2sin^2(\theta)p \Big),
\end{multline}
where $p = p(r)$ is the pressure, $\epsilon = \epsilon(r)$ is the energy density, and $d\Omega^2 = d \theta^2 + sin^2(\theta)d\phi^2$. Then from the conservation of energy-momentum $\left( \nabla_\mu T^{\mu \nu} = 0  \right)$, we obtain the TOV equations
\begin{align}
    &\frac{dM(r)}{dr} = 4\pi r^2 \frac{\epsilon(r)}{c^2} \label{dmdr} \\ 
    &\frac{dp}{dr} = - \left( \epsilon(r) +p(r) \right) \frac{d\alpha(r)}{dr}\\
    &\frac{d\alpha(r)}{dr} = \frac{1}{c^2} \frac{Gc^2M(r) + 4\pi r^3 G p(r)}{r \left[rc^2-2GM(r)  \right] },
\end{align}
where $M(r)$ is the gravitational mass of the stellar interior as a function of radius, $r$. Since we are modeling neutron stars mixed with dark matter, we adopt the two-fluid formalism. The two-fluid formalism assumes that the two fluids do not interact, except through gravity, so each fluid satisfies its own separate conservation of energy-momentum equation. Conservation of energy-momentum for each separate fluid is equivalent to letting 
\begin{align}
    p(r) &= p_B(r) + p_{\chi}(r)\\
    \epsilon(r) &= \epsilon_B(r) + \epsilon_\chi(r),
\end{align}
where $\epsilon_{B}$ and $\epsilon_\chi$ are the baryonic matter and ADM energy densities respectively, and $p_B$ and $p_\chi$ are the pressures of baryonic matter and ADM respectively.  This choice yields the two-fluid TOV equations:
\begin{align}
     &\frac{d\alpha(r)}{dr} = \frac{1}{c^2} \frac{Gc^2M(r) + 4\pi r^3 G p(r)}{r \left[rc^2-2GM(r)  \right] } \label{dalphadr}\\
      &\frac{dp_B}{dr} = - \left( \epsilon_B +p_B \right) \frac{d\alpha(r)}{dr} \label{dpbdr}\\
       &\frac{dp_\chi}{dr} = - \left( \epsilon_\chi +p_\chi \right) \frac{d\alpha(r)}{dr} \label{dpchidr}\\
        &\frac{dM_B(r)}{dr} = 4\pi r^2 \frac{\epsilon_B(r)}{c^2} \\ 
         &\frac{dM_\chi(r)}{dr} = 4\pi r^2 \frac{\epsilon_\chi(r)}{c^2},
\end{align}
where $M_\chi(r)$ is the gravitational mass of ADM, $M_B$ is the gravitational mass of baryonic matter, and $M(r) = M_B(r)+M_\chi(r)$.

When studying ADM admixed neutron stars, it is helpful to define the ADM mass-fraction, $F_\chi$, which gives the relative amount of gravitational mass of ADM compared to the total gravitational mass of the admixed neutron star. The mass-fraction is defined as
\begin{equation}
    F_\chi = \frac{M_\chi(R_\chi)}{M_\chi(R_\chi)+M_B(R_B)} \label{F_chi},
\end{equation}
where $M_\chi(R_\chi)$ is the total accumulated ADM gravitational mass evaluated at the ADM core radius and $M_B(R_B)$ is the baryonic matter gravitational mass evaluated at the baryonic radius. The mass-fraction is important to the analysis of ADM in neutron stars because the ADM core distribution is dependent on $F_\chi$, as well as the ADM particle mass ($m_\chi$) and the effective self-interaction strength ($g_\chi/m_\phi$) \cite{Sagun_2022_Boson,Miao_2022}. 

\section{Calculating the Baryonic and ADM Equations of State}\label{eosnumerical}
We expect the neutron star core to potentially have two components: baryonic and ADM. This necessitates accounting for two EoSs. Traditionally, the baryonic matter EoS is represented by one or more of the many tabulated EoSs \cite{Ellis2018,Nelson_2018,Kain2021,Husain2021,Bell2020,Das_2021} or the $\sigma-\omega$ model for the nuclear EoS \cite{Das2020,Apran_2019}. We use the general parametrized piece-wise polytropic models of \cite{Raaijmakers2019,Raaijmakers2020,Raaijmakers2021} with three polytropes divided by varying transition densities \citep[see][]{Hebeler13}. Since neutron stars have central densities above the nuclear saturation density $n_0 = 0.16 \,fm^{-3}$, uncertainties in the EoS become significant due the higher order effects of nucleon interactions at high densities. These uncertainties in the EoS can be captured by calculating the nucleon interactions within chiral effective field theory \citep{Hebeler10, Hebeler13}. For the EoS model, we employ a polytropic fit to these calculations between $0.5n_0$ and $1.1n_0$, which we connect at lower densities to the Baym-Pethick-Sutherland (BPS) crust EoS \citep{Baym71}. At higher densities, the formalism of the chiral effective field theory breaks down and we connect to the piece-wise polytropic parameterization.

The second component that we need to solve the two-fluid TOV equation is the bosonic ADM EoS of the \cite{Nelson_2018} ADM model. A derivation of the bosonic ADM EoS calculation is detailed in Appendix \ref{appendix A}, which finds
\begin{align}
    &\epsilon_\chi (r) = m_\chi c^2 n_\chi + \frac{g_\chi^2}{2 m_\phi^2} \frac{\hbar^3}{c} n_\chi^2 \label{ADM_EoS1}\\
    &p_\chi (r) = -\epsilon_\chi + \Tilde{\mu}_\chi n_\chi = \frac{g_\chi^2}{2 m_\phi^2} \frac{\hbar^3}{c} n_\chi^2 \label{ADM_EoS2}\\
    &\Tilde{\mu}_\chi (r) = \frac{\partial \epsilon_\chi}{\partial n_\chi},
\end{align}
where $n_\chi$ is the number density of ADM and $\Tilde{\mu}_\chi$ is the ADM chemical potential in the local Lorentz frame. We note that the overall placement of the constants $\hbar$ and $c$ in the last term of Eq.~(\ref{ADM_EoS1}) appear to be different from the same term in \cite{Miao_2022}. However, we implement $g_\chi/m_\phi$ with units of $c^2/\mathrm{MeV}$, which results in an identical placement of the constants.

We use a numerical algorithm to compute the mass-radius relation of constant $F_\chi$ [see Eq.(~\ref{F_chi})] using the two-fluid TOV equation and the EoS of both baryonic matter and ADM. As discussed in Sec.~\ref{twofluidTOV}, the mass-radius relation of a purely baryonic neutron star is found by varying its central energy density. However, in dark matter admixed neutron stars, the central energy densities of both ADM and baryonic matter can be varied to obtain a mass-radius relation. To maintain a constant $F_\chi$ along the mass-radius relation, the central energy densities of ADM and baryonic matter need to be varied simultaneously. For each variation in baryonic central energy density, there is an ADM central energy density that will result in the desired constant mass-fraction, $F_{\chi,desired}$. To find the ADM central energy density that results in $F_{\chi,desired}$, we iteratively compute the difference between $F_{\chi,desired}$ and test mass-fractions using the \textit{Scipy.optimize.brenth} python root-finding algorithm. The \textit{Scipy.optimize.brenth} algorithm works by using Brent's method \citep[see][]{Bus1975} to test ADM central energy densities until it finds a ``test'' mass-fraction that results in a difference of zero. The ADM central energy density that corresponds to this mass-fraction is saved. We repeat this process until each baryonic central energy density has a corresponding ADM central energy density which yields $F_{\chi,desired}$. 

\begin{figure}
    \centering
    \includegraphics[width = \columnwidth,scale = 1.5]{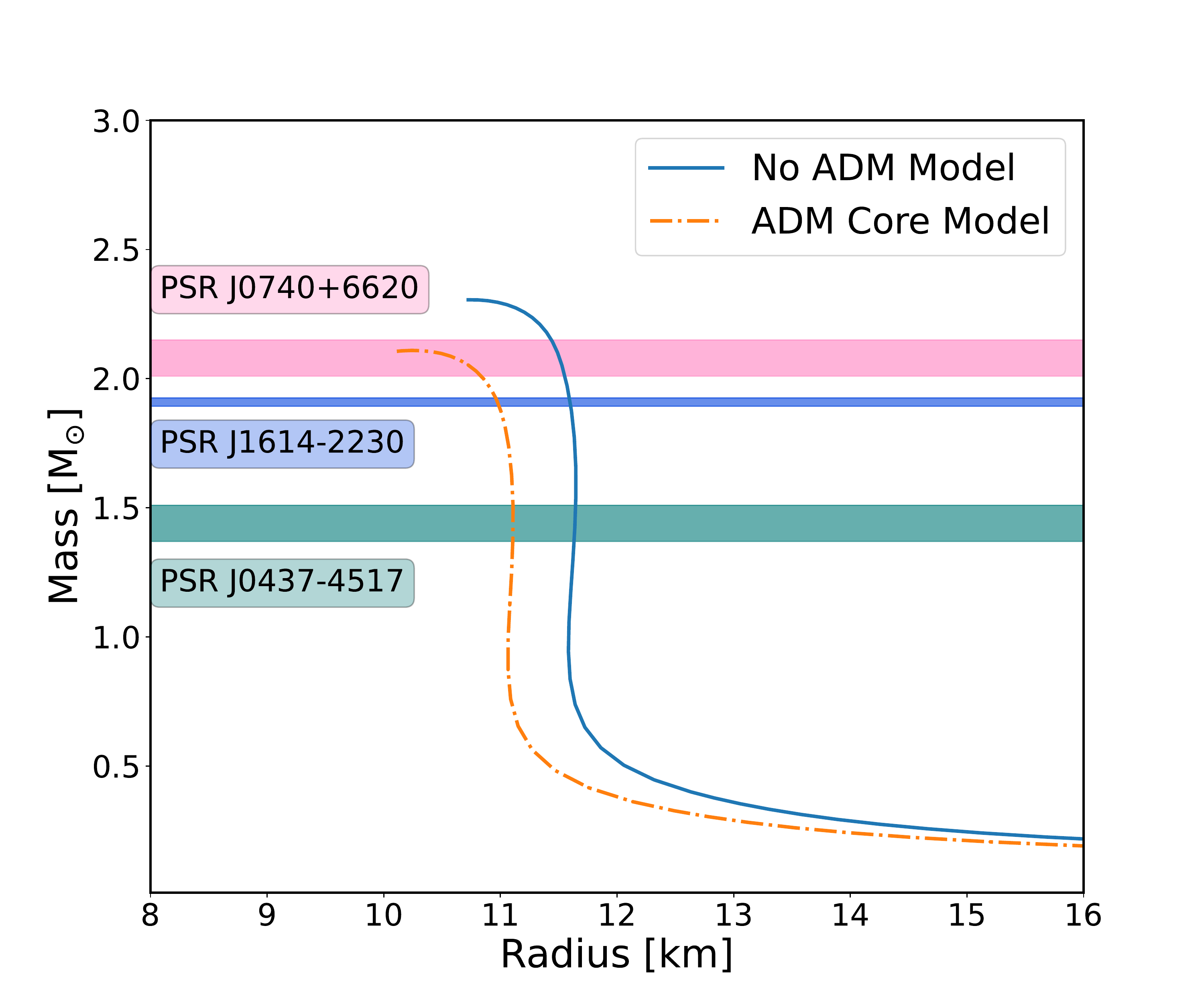}
    \caption{The mass-radius relations of the ground truth models described in Sec.~\ref{Varying BM EoS}. The first model assumes an ADM component to be present in neutron stars (dashed-dotted orange), while the second model does not have ADM in neutron stars (solid blue). The two relations are overlaid with the 68\% credible interval mass measurements of NICER sources PSR J0740+6620, PSR J1614-2230, and PSR J0437-4715.}
    \label{fig1}
\end{figure}

The mass-radius relation of constant $F_\chi$ is formed by iteratively repeating the procedure discussed previously for each considered baryonic matter central energy density. An example of the algorithm's result is shown in Fig.~\ref{fig1}.

\section{\textit{Future} and \textit{Future-X} Scenarios}\label{Best Case NICER and STROBE-X Scenarios}
 We will now show how we can constrain underlying physics using PPM-derived mass-radius measurements. We explain our inference framework, constraints on the ADM parameter space, simulated source selection, and finally, how we conducted the Bayesian analysis for two realistic hypothetical simulated scenarios. For both the \textit{Future} and \textit{Future-X} scenarios, we perform a Bayesian parameter estimation using a fixed baryonic matter EoS and with a varying baryonic matter EoS.

\subsection{Inference framework}\label{Inference Framework}
In order to infer the ADM parameters $m_\chi$, $g_\chi/m_\phi$, and $F_\chi$, we employ nested sampling, following the framework outlined in \cite{Greif19, Raaijmakers2019, Raaijmakers2020, Raaijmakers2021}. Letting $\boldsymbol{\theta}$ be a vector containing all baryonic and ADM EoS parameters, and $\boldsymbol{\epsilon_c}$ a vector containing the central baryonic and dark matter energy densities of observed stars, we can use Bayes' theorem to write the posterior distribution on $\boldsymbol{\theta}$ and $\boldsymbol{\epsilon_c}$ as\\
\begin{align}
    p(\boldsymbol{\theta}, \boldsymbol{\epsilon_c} |\mathbf{d})\nonumber &\propto p(\boldsymbol{\theta}) p(\boldsymbol{\epsilon_c}|\boldsymbol{\theta}) ~p(\mathbf{d} |\boldsymbol{\theta}, \boldsymbol{\epsilon_c}) \\
    & \propto p(\boldsymbol{\theta})  p(\boldsymbol{\epsilon_c}|\boldsymbol{\theta}) p\boldsymbol{(}\mathbf{d}|\mathbf{M}(\boldsymbol{\theta},\boldsymbol{\epsilon_c}), \mathbf{R}(\boldsymbol{\theta},\boldsymbol{\epsilon_c})\boldsymbol{)} \label{eq22}, 
\end{align}
where $\mathbf{d}$ is the vector containing the masses and radii of the sources from each scenario, $\mathbf{M}(\boldsymbol{\theta},\boldsymbol{\epsilon_c})$ is the mass of a produced admixed neutron star, and $\mathbf{R}(\boldsymbol{\theta},\boldsymbol{\epsilon_c})$ is the radius of the admixed neutron star. Equation~(\ref{eq22}) shows that the posterior distribution of the baryonic and ADM parameters is directly proportional to the product of the priors $p(\boldsymbol{\theta})  p(\boldsymbol{\epsilon_c}|\boldsymbol{\theta})$ and likelihood evaluation of the sources given the admixed neutron star mass and radii produced by the baryonic and ADM parameters.
If we furthermore equate the likelihood function $p\boldsymbol{(}\mathbf{d}|\mathbf{M}(\boldsymbol{\theta},\boldsymbol{\epsilon_c}), \mathbf{R}(\boldsymbol{\theta},\boldsymbol{\epsilon_c})\boldsymbol{)}$ to the PPM-derived mass-radius posteriors, and assume each of these posteriors is independent of the other, we can write\\
\begin{align}
\label{eq:posterior}
    p(\boldsymbol{\theta}, \boldsymbol{\epsilon_c} |\mathbf{d}) \propto~& p(\boldsymbol{\theta}) p(\boldsymbol{\epsilon_c}|\boldsymbol{\theta}) \\ & \prod_{i} p(M_i, R_i ~|~d_{PPM,i}) \nonumber, 
\end{align}
where $d_{PPM,i}$ is an element in the $\mathbf{d}$ vector and $i$ runs over the number of stars for which PPM delivers the mass and radius. 

The inference methods of \cite{Greif19, Raaijmakers2019, Raaijmakers2020, Raaijmakers2021} sample over the central energy density. We instead sample over the mass-fraction as defined in Eq.~(\ref{F_chi}), i.e., we introduce $F_{\chi} = F_{\chi}(\boldsymbol{\theta}, \boldsymbol{\epsilon_{c,B}}, \boldsymbol{\epsilon_{c,ADM}})$ and write Eq.~(\ref{eq:posterior}) as 
\begin{align}
    p(\boldsymbol{\theta}, \boldsymbol{\epsilon_{c,B}}, \nonumber \boldsymbol{F_{\chi}} |\mathbf{d}) \propto~& p(\boldsymbol{\theta}) p(\boldsymbol{\epsilon_{c, B}}|\boldsymbol{\theta}) p( \boldsymbol{F_{\chi}} |\boldsymbol{\theta}, \boldsymbol{\epsilon_c}) \\ & \prod_{i} p(M_i, R_i ~|~d_{PPM,i}),
\end{align}
where $\boldsymbol{\epsilon_{c,B}}$ and $\boldsymbol{\epsilon_{c,ADM}}$ are the central energy densities of baryonic matter and ADM, respectively. We sample over the mass-fraction because our mass-radius algorithm is structured such that the dark matter energy density is dependent on the mass-fraction.

\subsection{Bosonic ADM EoS constraints} \label{ADM constraints}
Here we define the prior space of each bosonic ADM parameter we wish to study, namely $m_\chi$, $g_\chi/m_\phi$, and $F_\chi$.  These independent parameters fully characterize our bosonic dark matter model, providing a parameter space to explore.  

To capture the bosonic ADM particle mass prior space, we consider the physical constraints on the parameter space placed by \cite{Kouvaris2011_pt2}, \cite{Bramante_2013}, and \cite{Bell_2013}. \cite{Kouvaris2011_pt2} showed that in order to safely avoid the ADM particles from escaping the neutron star,
\begin{equation}
     m_\chi \geq 10^{-2} \, \mathrm{MeV}.
\end{equation}
This provides a lower bound on the ADM particle mass prior space. The upper bound on the ADM particle mass can be found by considering observations from old neutron stars and the formation of a black holes in neutron stars from bosonic ADM core collapse \cite{Bramante_2013,Bell_2013}. According to \cite{Bramante_2013}, the upper bound on the ADM particle mass is $\approx 10^{5}-10^{7}$ GeV depending on the self-interaction of the ADM particles. In order to evade bosonic ADM core collapse due to an inadequate self-repulsion strength, we adopt a restrictive upper bound on $m_\chi$ to be
\begin{equation}
    m_\chi \leq 10^8 \, \mathrm{MeV}.
\end{equation}

 Although the bosonic ADM particle mass prior space is physically constrained, the repulsive self-interaction strength is not. To ensure the self-repulsion prior space has a finite size, we adopt a range that captures the interaction strengths used in \cite{Nelson_2018}
\begin{equation}
    10^{-2} \, \leq \frac{g_\chi}{m_\phi/\mathrm{MeV}} \leq 10^3 \,.
\end{equation}
The repulsive self-interaction strength is critical for our bosonic dark matter model because, in the absence of any self-repulsion, the dark matter would quickly form a black hole and destroy the neutron star \cite{Nelson_2018,Kouvaris_2010,Kouvaris2015,Bell_2013}.

The final ADM parameter of our bosonic ADM model is the mass-fraction, which is constrained by prior work to be $F_\chi \leq 5 \%$ \cite{Sagun_2022_Boson}. This result was found by computing $F_\chi(m_\chi)$ at a fixed self-interaction strength for a singular baryonic EoS. The results of the function were compared to the $1.4 \Msun$ neutron star tidal deformability constraint of GW170817 from \cite{GW170817eos} and the requirement that a mass-radius relation must have a maximum mass of at least $2 \Msun$. Since we vary the baryonic EoS, we adopt the upper bound of  
 \begin{equation}
     F_\chi \leq 20 \%,
 \end{equation}
 to account for the changes in the mass-fraction constraint from \cite{Sagun_2022_Boson}.
Based on the constraints mentioned above, we set the priors of the ADM parameter space to be
 \begin{align}
     &\log_{10}(m_\chi/\mathrm{MeV}) \in [-2,8]\\
     & \log_{10}\big(\frac{g_\chi}{m_\phi/\mathrm{MeV}}\big) \in [-2,3] \\
     & F_\chi \in [0,20] \, \% .
 \end{align}
In each interval above, we uniformly sample each parameter without considering the formation of a dark matter core or halo. After sampling, we find all the combinations of $m_\chi,\, g_\chi/m_\phi, \, \mathrm{and} \, F_\chi$ that result in a halo configuration and give each a zero probability density, which leaves only the ADM core configurations in the prior space. This choice results in the prior corner plots of Fig.~\ref{fig2} displaying a nonuniform prior for each bosonic ADM parameter.
\begin{figure}
\centering
\begin{subfigure}[b]{0.5\textwidth}
   \includegraphics[width=\textwidth,scale = 1.5]{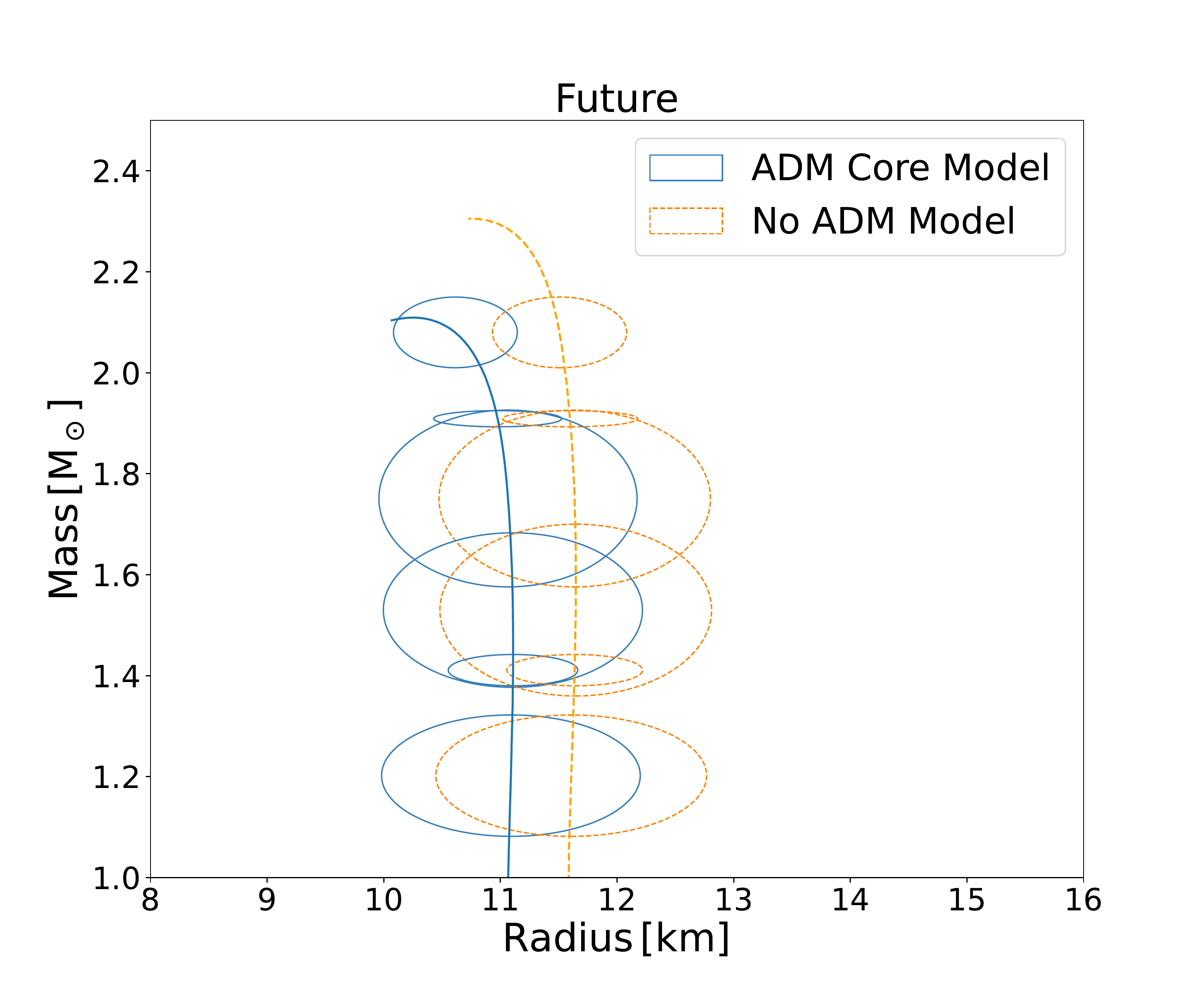}
   \caption{}
   \label{fig:Nr1} 
\end{subfigure}

\begin{subfigure}[b]{0.5\textwidth}
   \includegraphics[width=\textwidth,scale = 1.5]{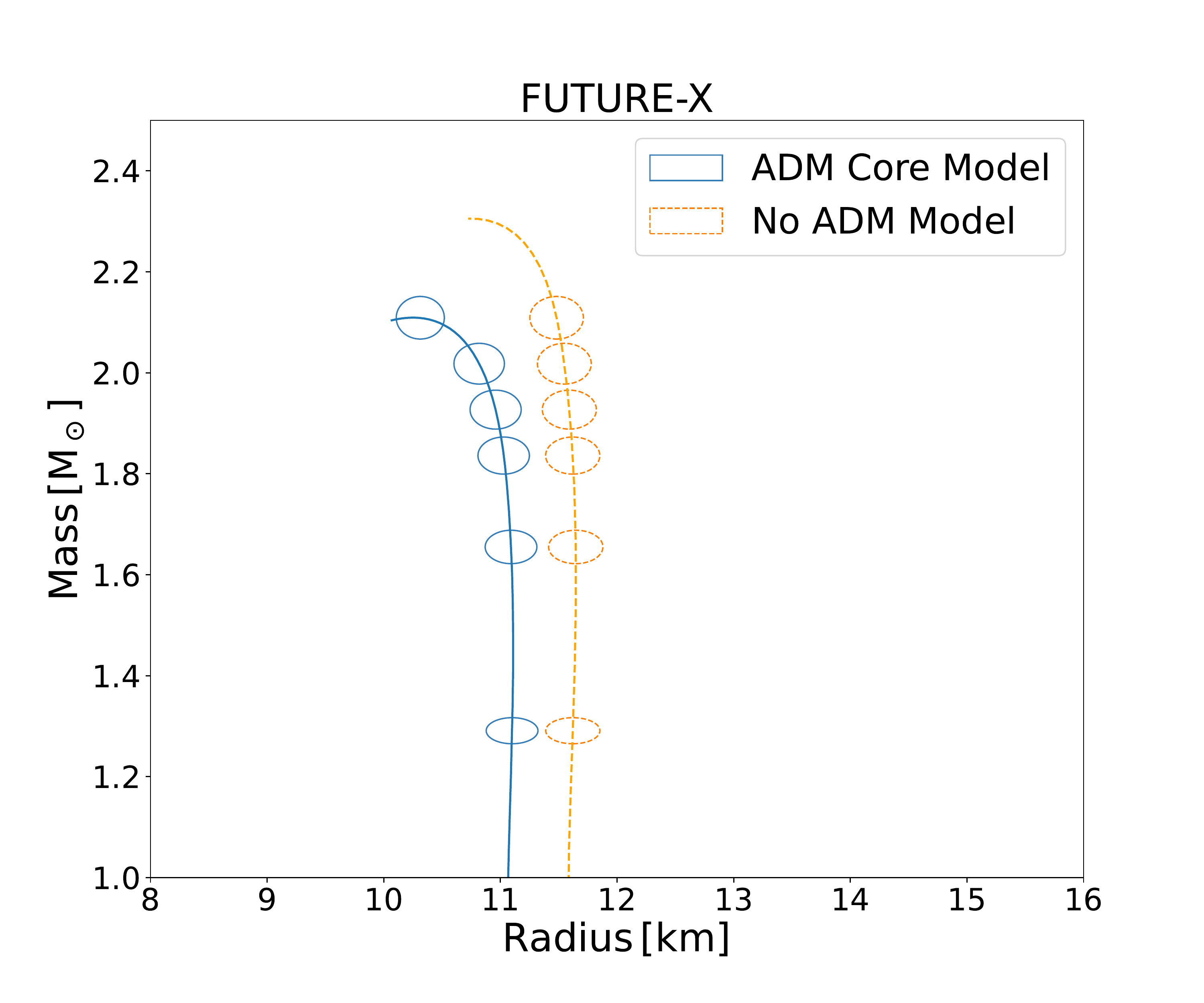}
   \caption{}
   \label{fig:Nr2}
\end{subfigure}
\caption{(a) Uncertainty ellipses from the 1 sigma level of a 2-D Gaussian for each of the synthetic \textit{Future} sources calculated from both ground truth models. (b) Same as in (a), but for the 6 \textit{Future-X} sources. For both panels, the corresponding ground models are overlaid on the appropriate ellipses.}
\label{figure_ellipses}
\end{figure}
 \begin{figure*}
\centering
\begin{subfigure}{.5\textwidth}
  \centering
  \includegraphics[width=\textwidth,scale = 1.5]{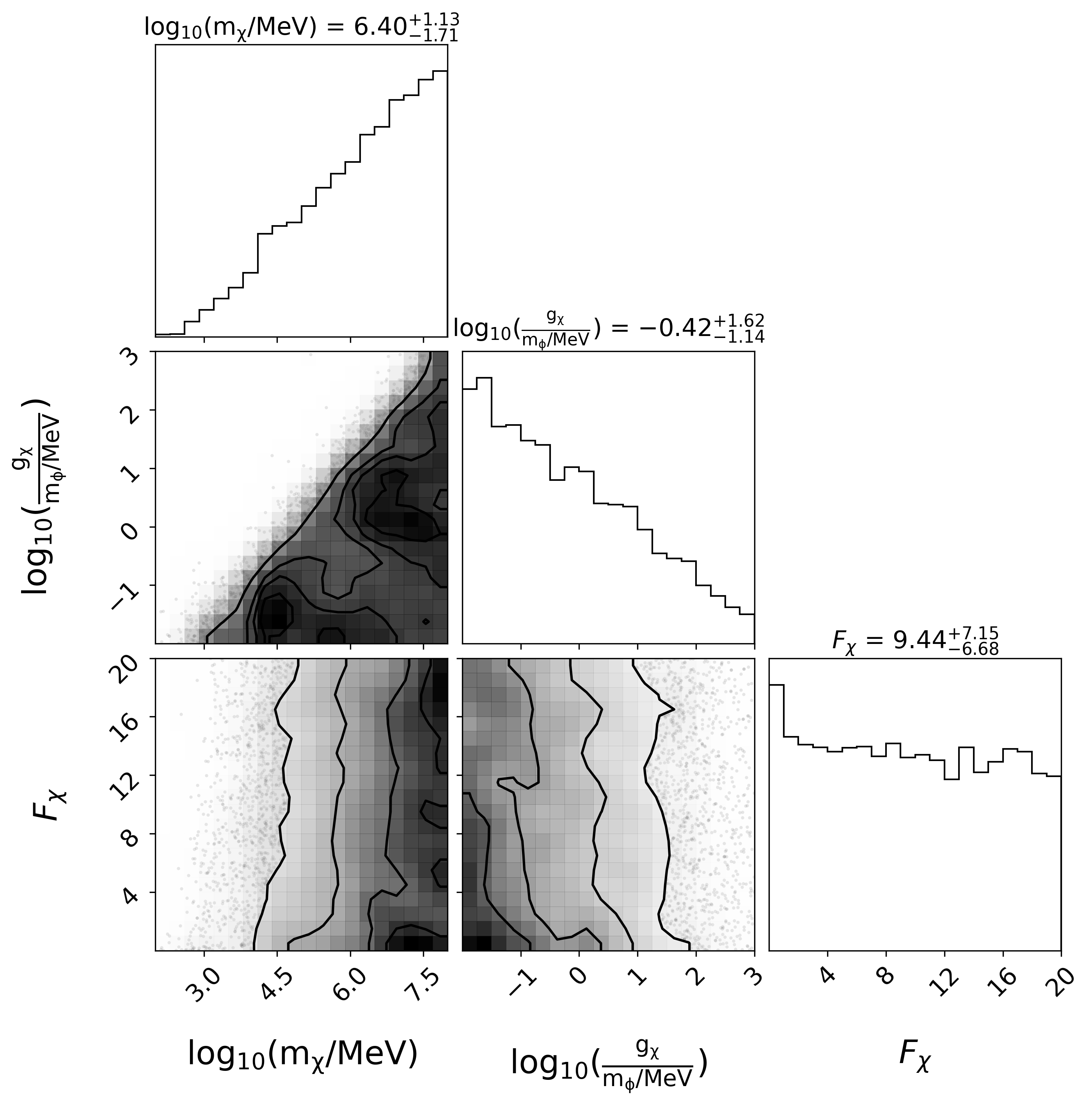}
  \label{fig2:sub1}
\end{subfigure}%
\begin{subfigure}{.5\textwidth}
  \centering
  \includegraphics[width=\textwidth]{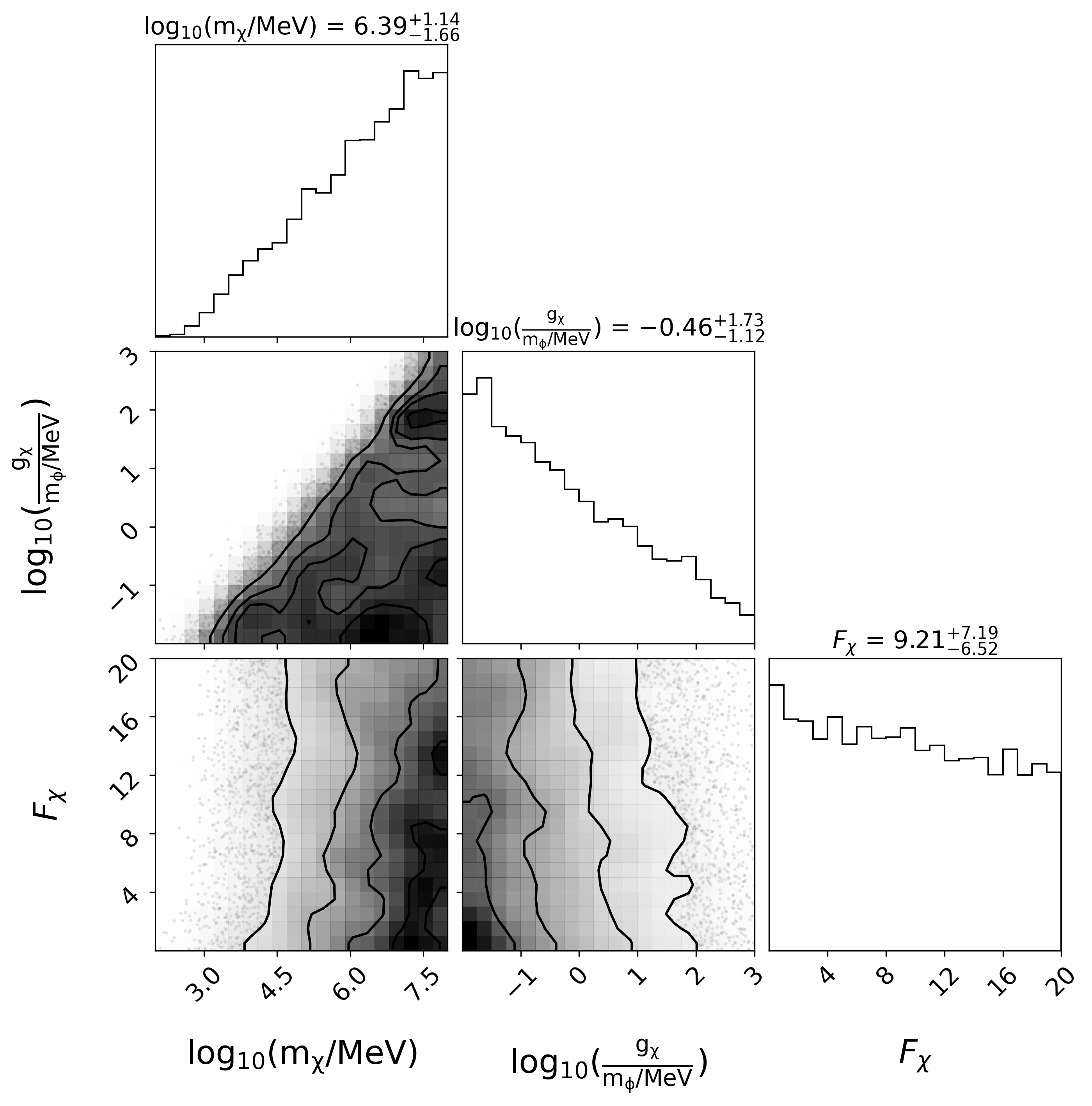}
  \label{fig2:sub2}
\end{subfigure}
\caption{(Left panel) \textit{Future} prior corner plot of the ADM parameters; (Right panel) The \textit{Future-X} prior corner plots of the ADM parameters. Both corner plots were generated using the corner python library \cite{corner}. For each panel, the ADM particle mass, effective self-interaction strength, and mass-fraction are plotted against each other, where the dark shaded regions represent a higher prior probability and lighter shaded regions represent a lower prior probability. The solid lines represent contours of constant probability, where we show the 0.5, 1, 1.5, 2 $\sigma$ level contours. When the combination of parameters is a parameter plotted against itself, we have a 1-D prior histogram. For both panels, the figure titles on the diagonal show the median value with the 0.16 and 0.84 fractional quantiles. In both panels, we see large regions of no shading in all 2-D prior density plots.}
\label{fig2}
\end{figure*}
\subsection{Source selection}\label{sources}
 Here we introduce the sources used in our Bayesian analysis for the \textit{Future} and \textit{Future-X} scenarios, respectively. For the \textit{Future} scenario, three of the six sources are assumed to have an \textit{a priori} known mass obtained from radio pulsar timing, corresponding to the NICER targets PSR J0740$+$6620 \cite{Fonseca_2021}, PSR J1614$-$2230 \cite{Zaven_2018}, and PSR J0437$-$4715 \cite{Reardon_2016}. For each of these known sources we assume a $5\%$ uncertainty in radius. Note that the current uncertainty on the radius of PSR J0740+6620 reported by NICER is closer to the $\sim 10$\% level, so achieving $5$\% is very much a stretch goal even given NICER's extended mission lifetime to 2025. For the other three sources masses were chosen between $1.2-2.1 \, \Msun$, and we assume a $10\%$ uncertainty in mass and radius, close to what NICER has already delivered for the one source (PSR J0030+0451) where no independent mass measurement is available. 
 
 For the \textit{Future-X} case we again assume six sources, but this time with a $2\%$ mass and radius uncertainty and selected in the range of $1.2-2.2 \, \Msun$. This scenario improves on what is possible with NICER and is something we anticipate being achievable with a large-area x-ray telescope operating for several years \citep{Watts2016,Ray2019,extp}. For STROBE-X, we expect to be able to make mass-radius measurements for $\approx 15$ sources using PPM at the $\pm 5\%$ uncertainty level \cite{Ray2019}. However, with longer observations, we expect STROBE-X to provide approximately six mass-radius measurements at the two percent uncertainty level which will likely provide the strongest EoS constraints. 

\begin{figure*}
\centering
\begin{subfigure}{.5\textwidth}
  \centering
  \includegraphics[width=\textwidth]{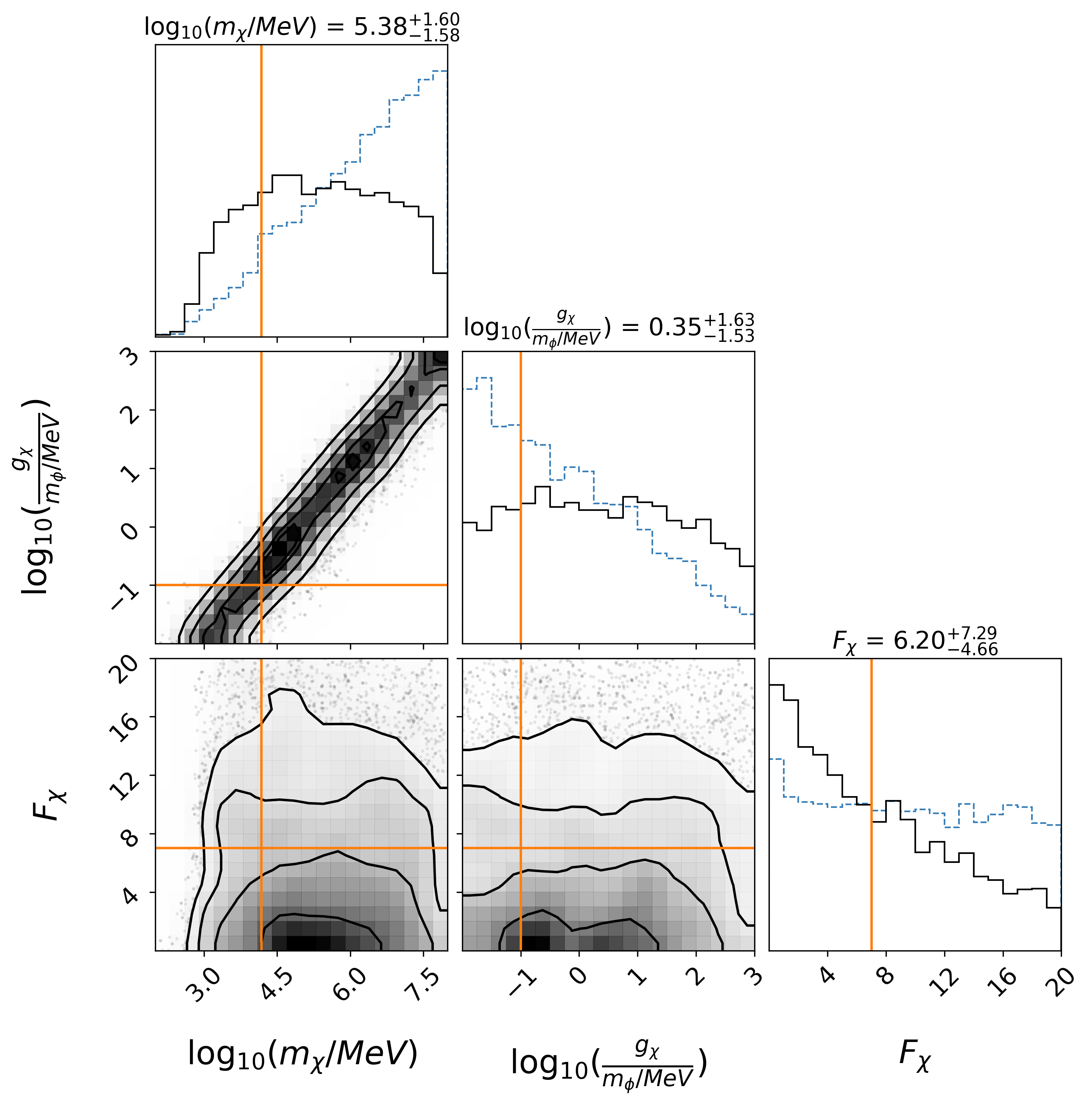}
  \label{fig3:sub1}
\end{subfigure}%
\begin{subfigure}{.5\textwidth}
  \centering
  \includegraphics[width=\textwidth]{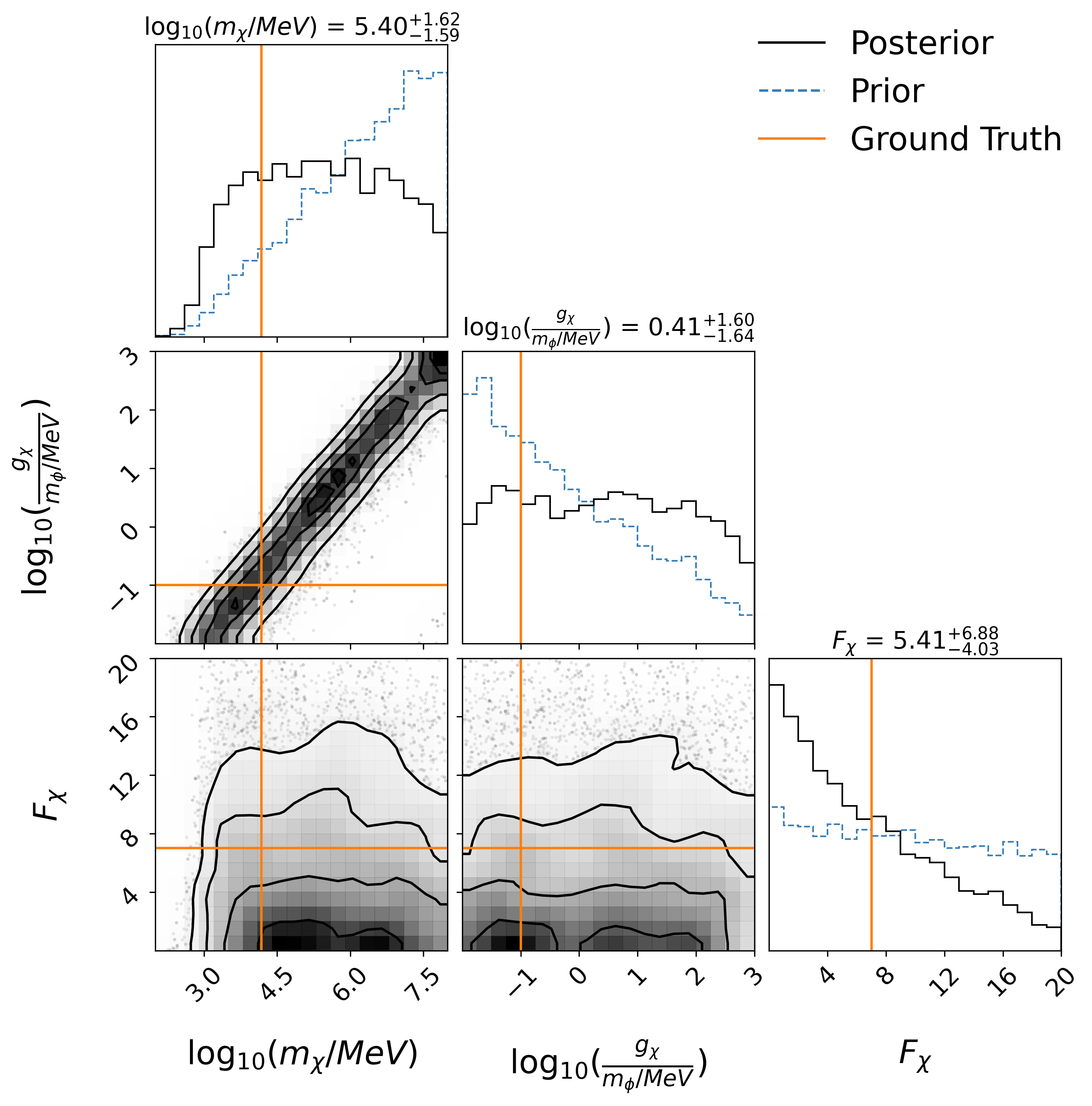}
  \label{fig3:sub2}
\end{subfigure}
\begin{subfigure}{.5\textwidth}
  \centering
  \includegraphics[width=\textwidth]{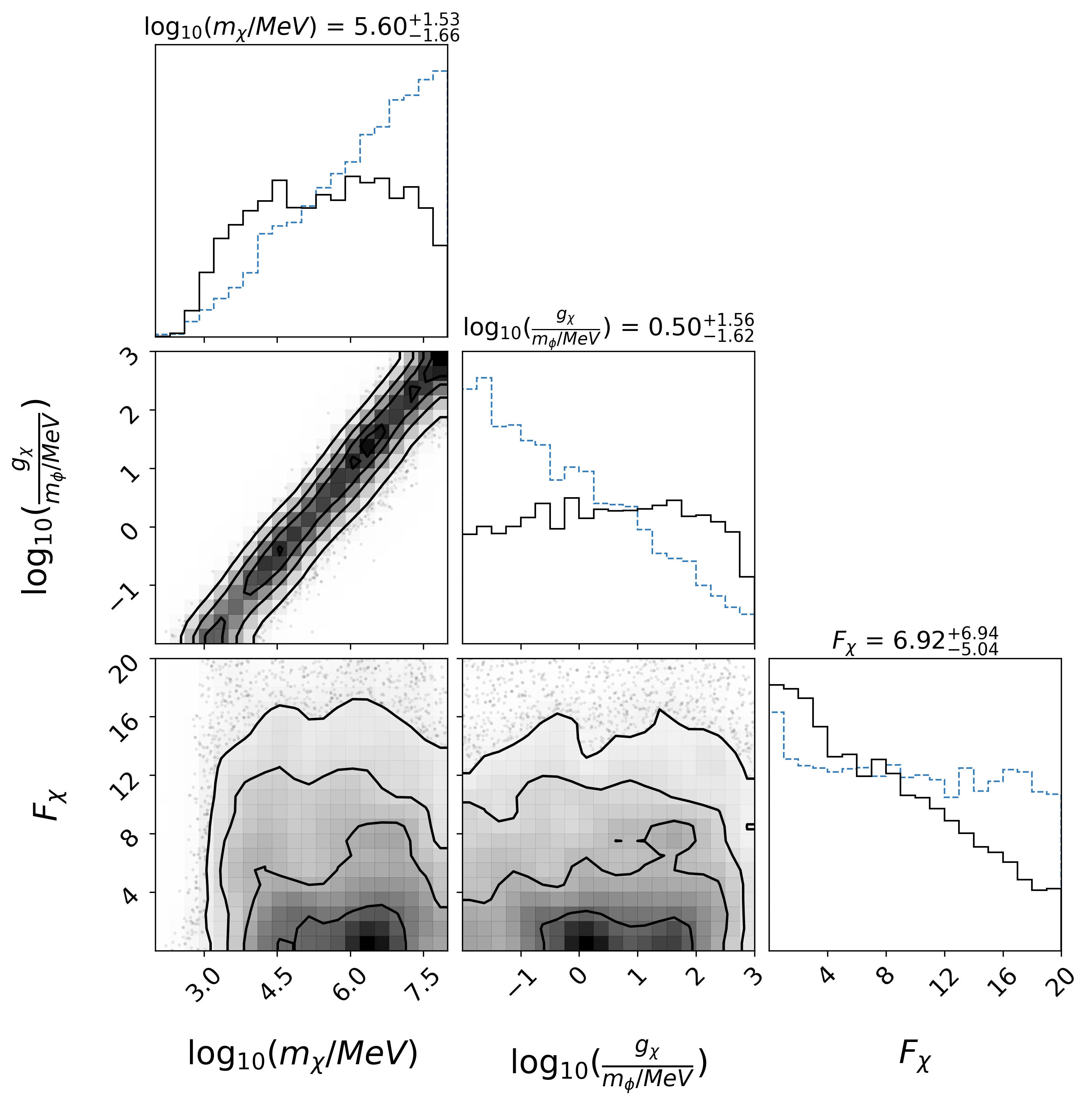}
  \label{fig3:sub3}
\end{subfigure}%
\begin{subfigure}{.5\textwidth}
  \centering
  \includegraphics[width=\textwidth]{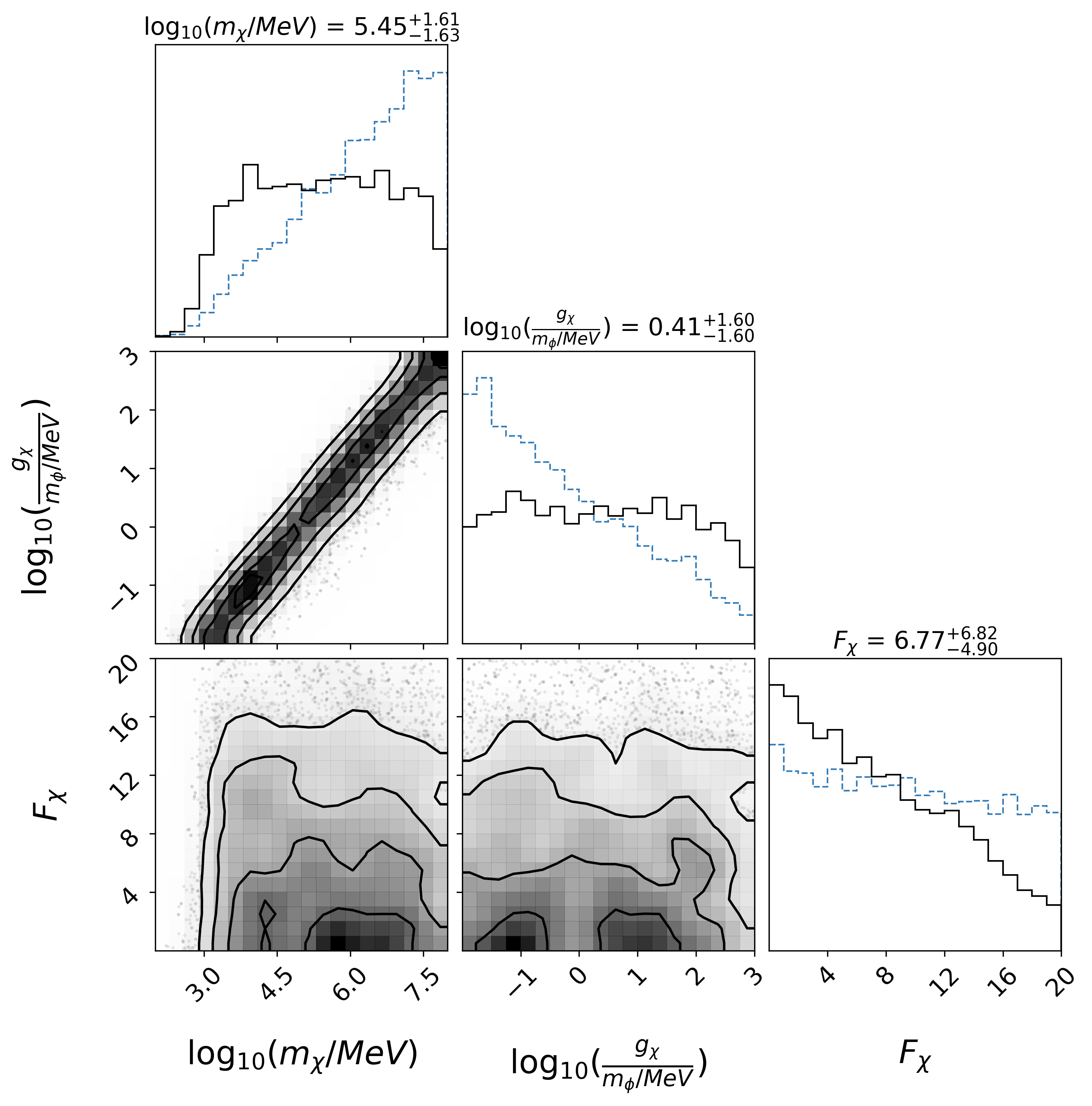}
  \label{fig3:sub4}
\end{subfigure}
\caption{Left two panels: \textit{Future} scenario with varying baryonic matter EoS posterior corner plot; Top Left: the ``ADM core'' ground truth model; Bottom Left: the ``No ADM'' ground truth model. (Right two panels) \textit{Future-X} scenario with varying baryonic matter EoS corner plot; Top Right: the ``ADM core'' ground truth model; Bottom Right: the ``No ADM'' ground truth model. For the top 2 panels, the solid orange lines represent the ground truth values of ADM values for the ``ADM core'' model. For all quadrants, the dashed blue lines represent the 1-D priors shown in Fig.~\ref{fig2} for the \textit{Future} and \textit{Future-X} cases, respectively. The solid black lines represent the posteriors of each case. In all panels, the ratio of $\mathrm{log_{10}}(g_\chi/(m_\phi/\mathrm{MeV}))$ and $\mathrm{log_{10}}(m_\chi/\mathrm{MeV})$ is constrained; however, the 1-D posterior of $F_\chi$ appears to be independent of this.}
\label{fig3}
\end{figure*}
\begin{figure*}
\centering
\begin{subfigure}{.5\textwidth}
  \centering
  \includegraphics[width=\textwidth]{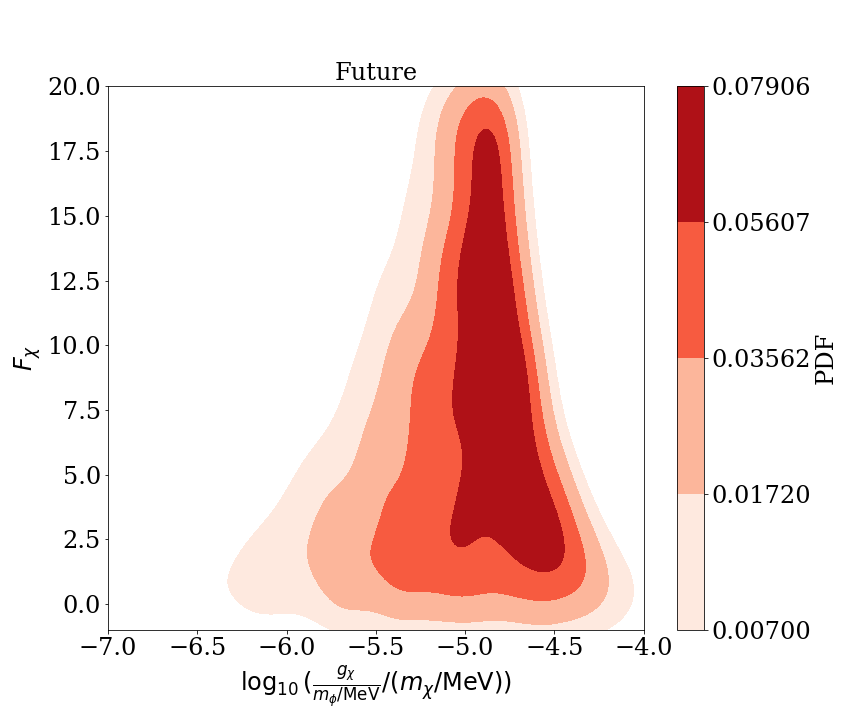}
  \label{fig3a:sub1}
\end{subfigure}%
\begin{subfigure}{.5\textwidth}
  \centering
  \includegraphics[width=\textwidth]{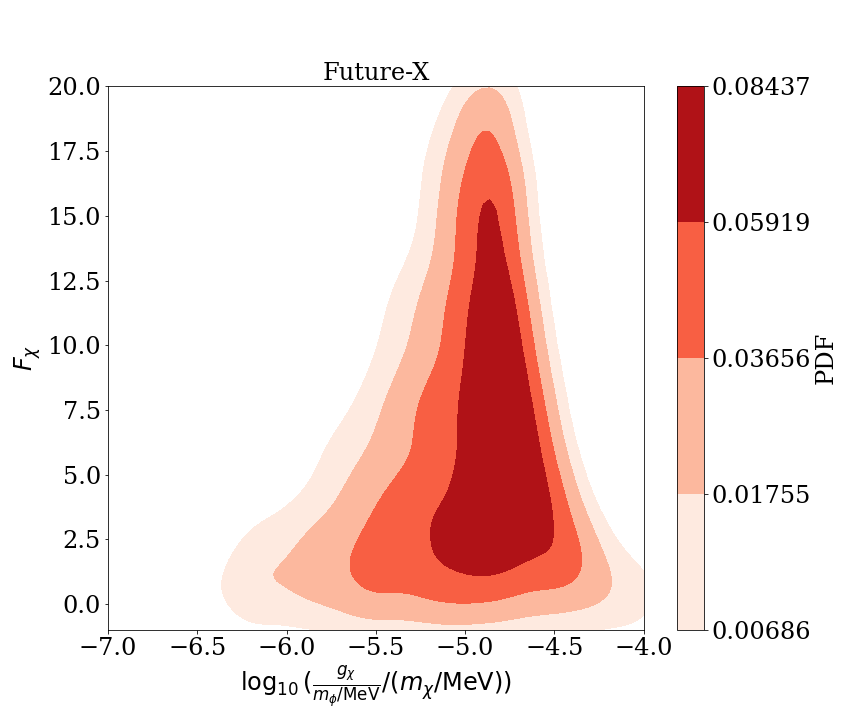}
  \label{fig3a:sub2}
\end{subfigure}
\begin{subfigure}{.5\textwidth}
  \centering
  \includegraphics[width=\textwidth]{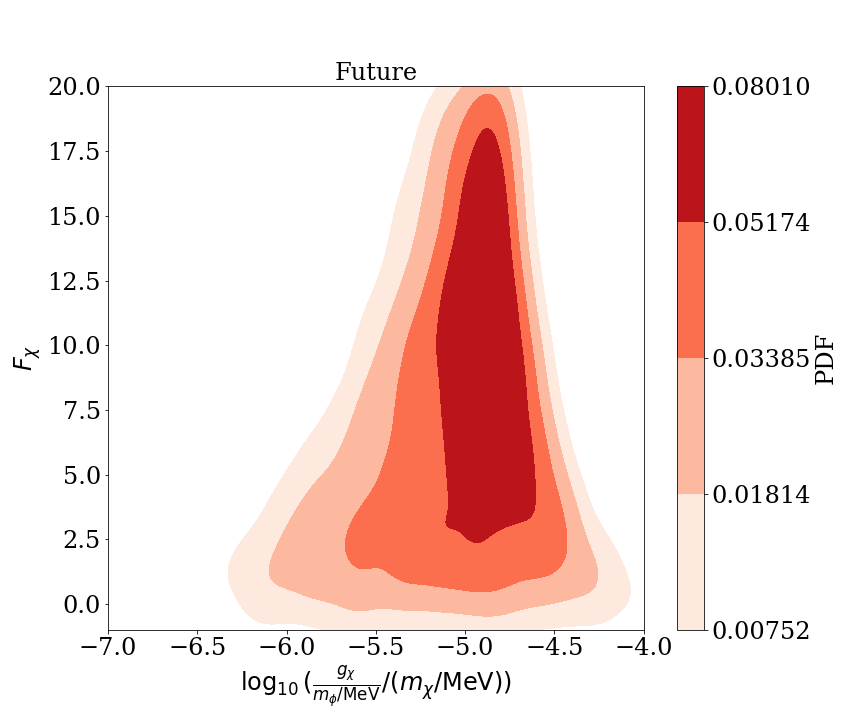}
  \label{fig3a:sub3}
\end{subfigure}%
\begin{subfigure}{.5\textwidth}
  \centering
  \includegraphics[width=\textwidth]{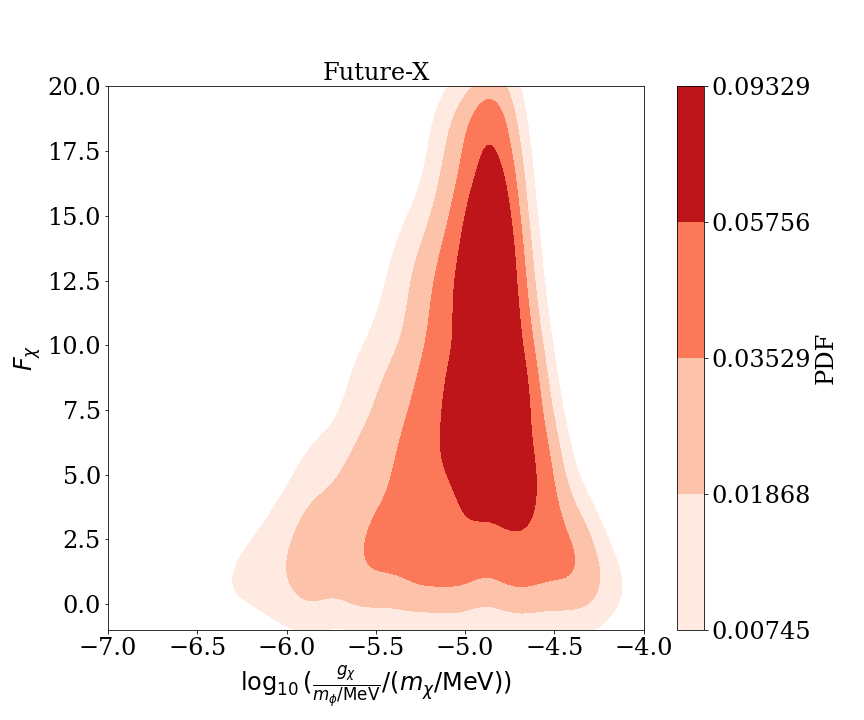}
  \label{fig3a:sub4}
\end{subfigure}
\caption{Left two panels: \textit{Future} scenario with posterior probability density function (PDF) plotted in the $F_\chi \, vs. \log_{10} \left(\frac{g_\chi}{(m_\phi/\mathrm{MeV})}/(m_\chi/\mathrm{MeV})\right)$ plane; Right two panels: \textit{Future-X} scenario with posterior PDF plotted in the $F_\chi \, vs. \log_{10} \left(\frac{g_\chi}{(m_\phi/\mathrm{MeV})}/(m_\chi/\mathrm{MeV})\right)$ plane. All contour plots were made with the seaborn python package \cite{Waskom2021}. For all quadrants, the posteriors and ground truth models follow the same flow as in Fig.~\ref{fig3}. The solid shaded regions represent the same contour levels described in Fig.~\ref{fig2}. In all panels, the posterior PDF contours widen along the $\log_{10} \left(\frac{g_\chi}{(m_\phi/\mathrm{MeV})}/(m_\chi/\mathrm{MeV})\right)$ axis for small $F_\chi$.}
\label{fig3a}
\end{figure*}
\begin{figure*}
\centering
  \includegraphics[width=\textwidth]{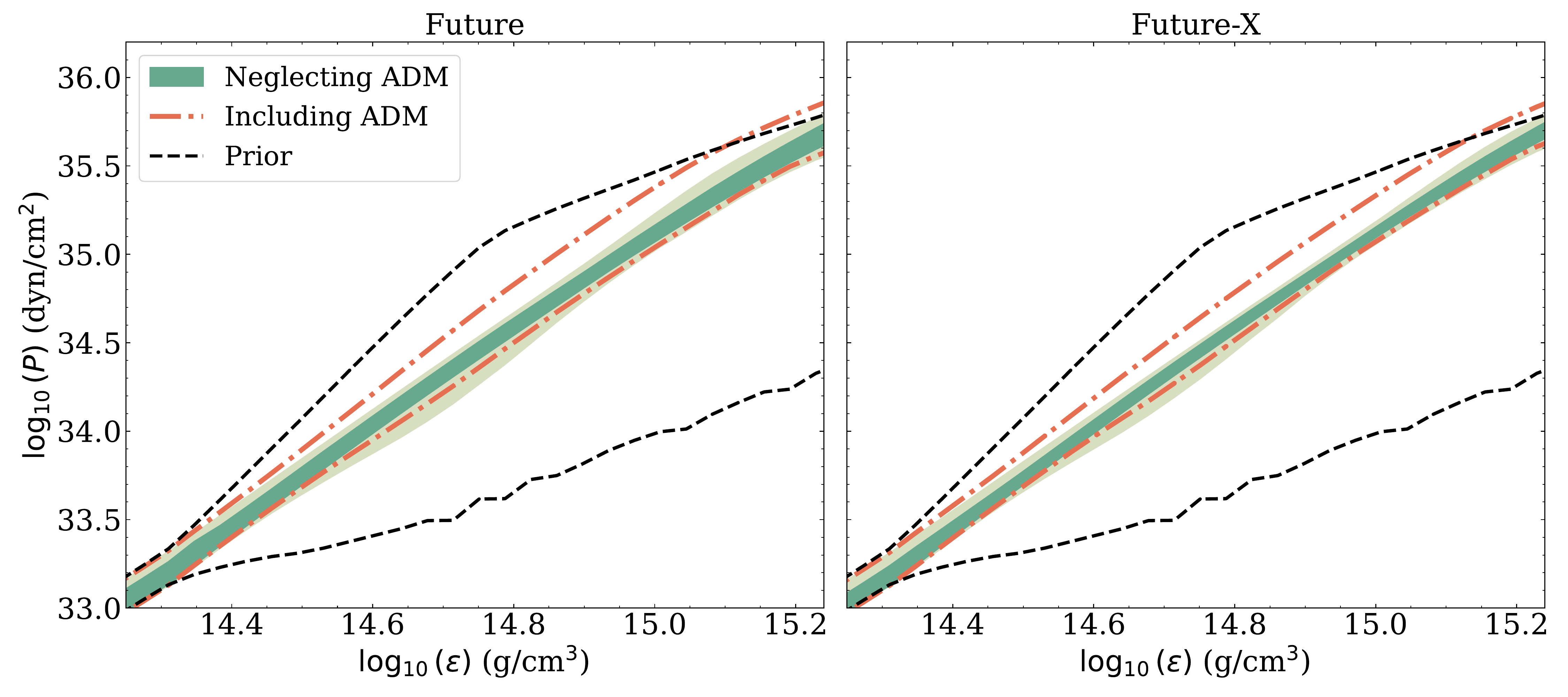}
 \includegraphics[width=\textwidth]{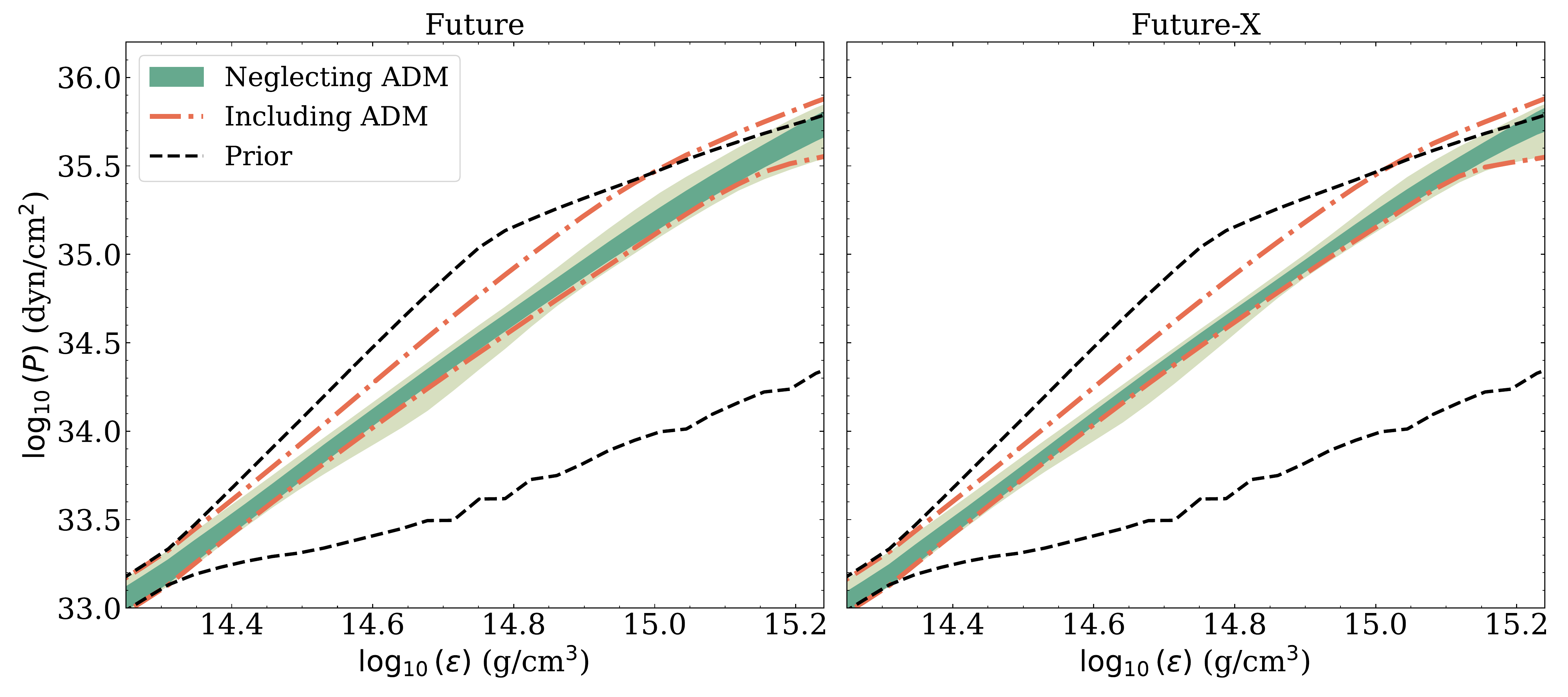}
\caption{Left two panels: \textit{Future} EoS plots with varying baryonic matter EoS; Right two panels: \textit{Future-X} EoS plot with varying baryonic matter EoS. Note the ground truth models used in each respective panels follows the same flow as in Fig.~\ref{fig3}. For all panels, the dashed black line represents the $95\%$ prior distribution, the light and dark green band represents the 68\% and 95\% confidence region of the posterior distribution when only varying the baryonic matter EoS, with both ground truth models. As a comparison, the dashed-dotted orange line is the $95\%$ confidence region when we additionally allow for ADM in the star, i.e., we also sample the ADM EoS parameters. In all panels, we see that including ADM in neutron stars widens the constraints on the baryonic EoS.}
\label{fig4}
\end{figure*}

\subsection{Best sase \textit{Future}/\textit{Future-X}: Varying baryonic EoS} \label{Varying BM EoS}
In both varying and fixed baryonic EoS cases, we calculate the radii of our sources from two ground truth models. The first ground truth model, which we call the ``ADM core" model, is described by the piecewise polytropic (PP) model in Sec.~\ref{eosnumerical} with an ADM core defined by the ADM parameters
\begin{align}
    m_\chi &= 15 \, \mathrm{GeV}\\
    \frac{g_\chi}{m_\phi/\mathrm{MeV}} &= 0.1\\
    F_\chi &= 7\%.
\end{align}
The second ground truth model, the ``No ADM" model, is the PP model mentioned in Sec.~\ref{eosnumerical} with $F_\chi = 0\%$, which simulates neutron star with no accumulated ADM, but ADM is still taken into account during the inference. The mass-radius relation of both ground truth models is shown in Fig.~\ref{fig1}. The uncertainty ellipses of each of the \textit{Future} and \textit{Future-X} sources are displayed in Fig.~\ref{figure_ellipses}. Notably, in \cite{Greif19}, the data points were scattered around the ground truth model, but not centered on the ground truth like our source ellipses are. This is more realistic. However, given that selecting a scatter model also introduces some arbitrariness -- for example the masses at which the scatter is smallest matters -- we have chosen not to do this at this stage. Effectively, this makes our current study a best case scenario.

The most conservative approach to a Bayesian analysis of neutron stars varies all parameters in the EoS model. Therefore, the most likely EoS, and by extension mass-radius relation, of baryonic matter and ADM can be inferred. In Fig.~\ref{fig2}, we show the prior distributions of the ADM parameters for the \textit{Future} and \textit{Future-X} scenarios. Both panels in Fig.~\ref{fig2} have similar regions in which there is no shading, e.g., the upper left region in the $\log_{10}\boldsymbol{(}g_\chi/(m_\phi/\mathrm{MeV})\boldsymbol{)} \, vs. \, \log_{10}(m_\chi/\mathrm{MeV})$ plots. This region is cut out because that part of the bosonic ADM parameter space is composed of halo configurations.

In Fig.~\ref{fig3} and \ref{fig3a}, we show the posterior distributions for both the \textit{Future} and \textit{Future-X} scenarios. The posterior of Fig.~\ref{fig3} shows that the ratio of $\log_{10}\boldsymbol{(}g_\chi/(m_\phi/\mathrm{MeV})\boldsymbol{)}$ and $\log_{10}(m_\chi/\mathrm{MeV})$ forms an identical stripe that runs diagonally across the parameter space in all four quadrants. Figure~\ref{fig3} also shows that, despite the ground truth value of 7\%, the 1-D $F_\chi$ posteriors of the ``ADM core" model favor low mass fractions in both scenarios. Additionally, the 1-D $F_\chi$ posteriors of all four plots are nearly identical and favor low mass-fractions. From the observation that the posteriors of both ground truth models are nearly identical, we conclude neither \textit{Future} nor \textit{Future-X} will be able to determine if bosonic ADM is present in neutron stars. Therefore, by extension, neither scenario will be able to constrain the bosonic ADM parameter space within the current uncertainties of the baryonic EoS. However, if the actual ADM mass-fraction is not small, we find that the ratio of $g_\chi/m_\phi$ and $m_\chi$ is well constrained because numerous mass-fraction values are sampled and not given a low likelihood based on the sources from either of the ground-truth models. This is clearly shown in Fig.~\ref{fig3a}, where 
 we observe that the PDF contours widen along the $\log_{10} \left(\frac{g_\chi}{(m_\phi/\mathrm{MeV})}/(m_\chi/\mathrm{MeV})\right)$ axis for low $F_\chi$.

In Fig.~\ref{fig4} we study our Bayesian inferences in the energy density-pressure plane. This has the advantage of highlighting constraints on the baryonic matter EoS that are not readily observable in the posterior corner plots. In all quadrants, we consider the EoS posteriors that only vary the baryonic EoS (`Neglecting ADM') and the EoS posteriors that additionally vary the ADM EoS (`Including ADM'). The `Including ADM' posteriors correspond to the posteriors shown in Fig.~\ref{fig3}. For both the \textit{Future} and \textit{Future-X} scenarios, Fig.~\ref{fig4} shows that the 95\% confidence interval of the `Including ADM' (orange dashed-dotted band) band is noticeably wider than the 95\% confidence interval of `Neglecting ADM' (light green band). In the \textit{Future scenario}, we calculate that the `including ADM' 95\% confidence interval is 12\% and 20\% wider than the `Neglecting ADM' 95\% confidence interval for the ``ADM core" and ``No ADM" models, respectively, at $\mathrm{log_{10}}(\epsilon \, \mathrm{cm^3/g}) = 14.71$. For \textit{Future-X}, we calculate that the `including ADM' 95\% confidence interval is 18\% and 25\% wider than the `Neglecting ADM' 95\% confidence interval for the ``ADM core" and ``No ADM" models, respectively, at $\mathrm{log_{10}}(\epsilon \, \mathrm{cm^3/g}) = 14.71$. Including ADM broadens the 95\% confidence interval because ADM cores decrease the radii of neutron stars, so the baryonic matter EoS is allowed to be more stiff and still be consistent with mass-radius measurements. A stiffer baryonic EoS implies the posterior constraints from all presently observed NICER pulsars, and the future sources of both NICER and STROBE-X, can be relaxed if the possibility of ADM is considered. Finally, this figure also demonstrates \textit{Future-X}'s ability to more tightly constrain the neutron star EoS due to the tighter confidence bands in both posteriors.

\begin{figure*}
\centering
\begin{subfigure}{.5\textwidth}
  \centering
  \includegraphics[width=\textwidth]{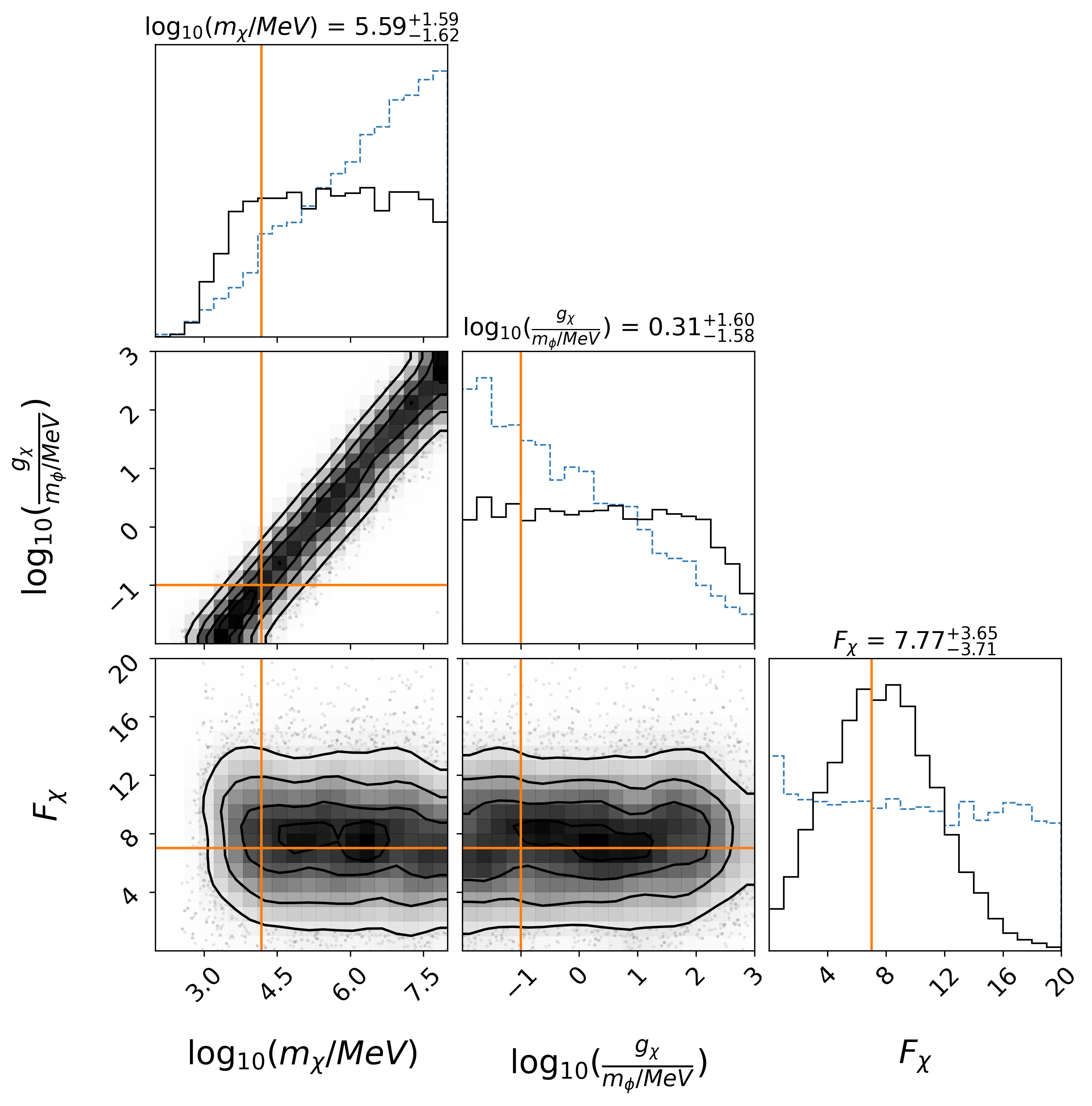}
  \label{fig5:sub1}
\end{subfigure}%
\begin{subfigure}{.5\textwidth}
  \centering
  \includegraphics[width=\textwidth]{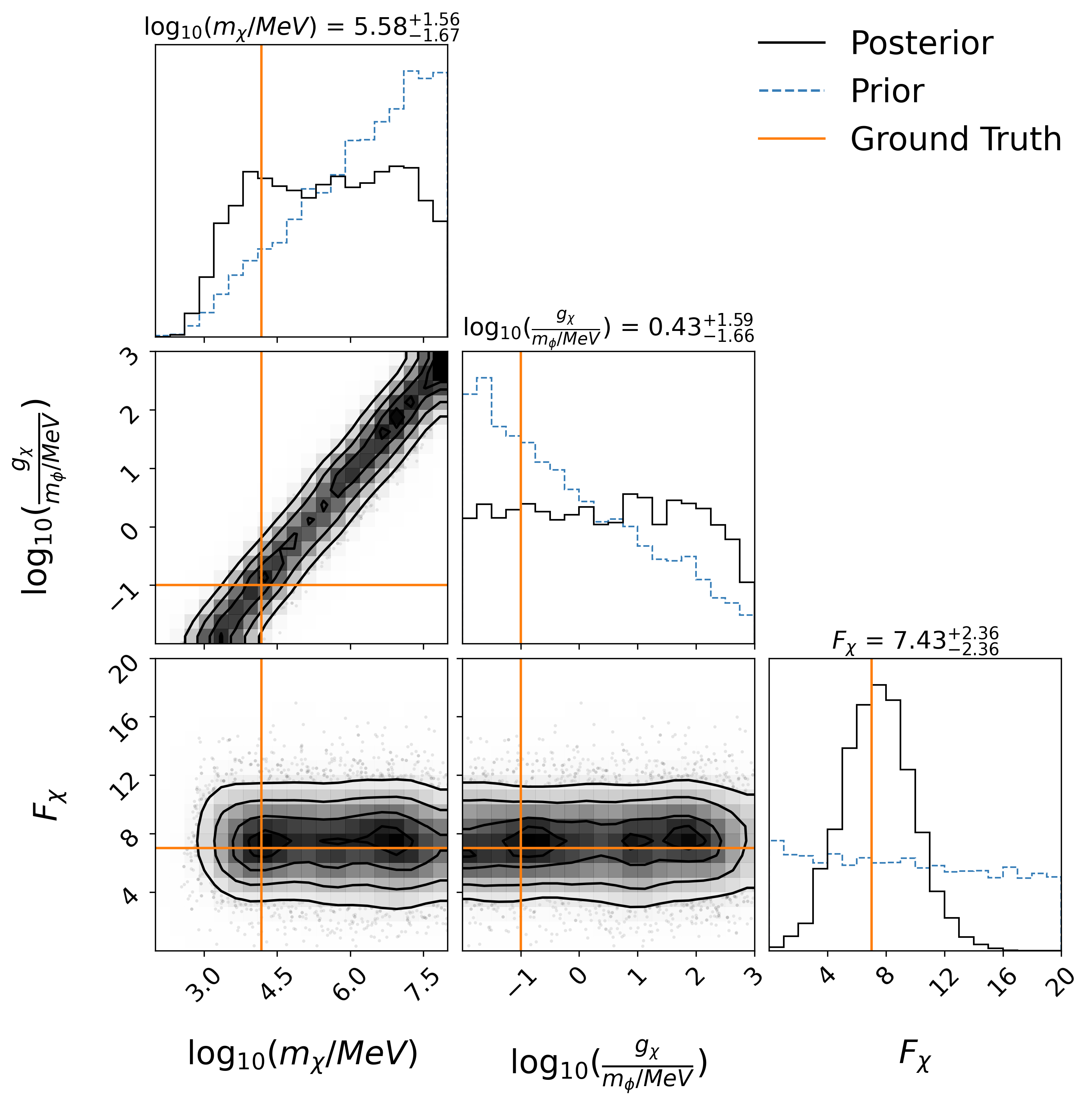}
  \label{fig5:sub2}
\end{subfigure}
\begin{subfigure}{.5\textwidth}
  \centering
  \includegraphics[width=\textwidth]{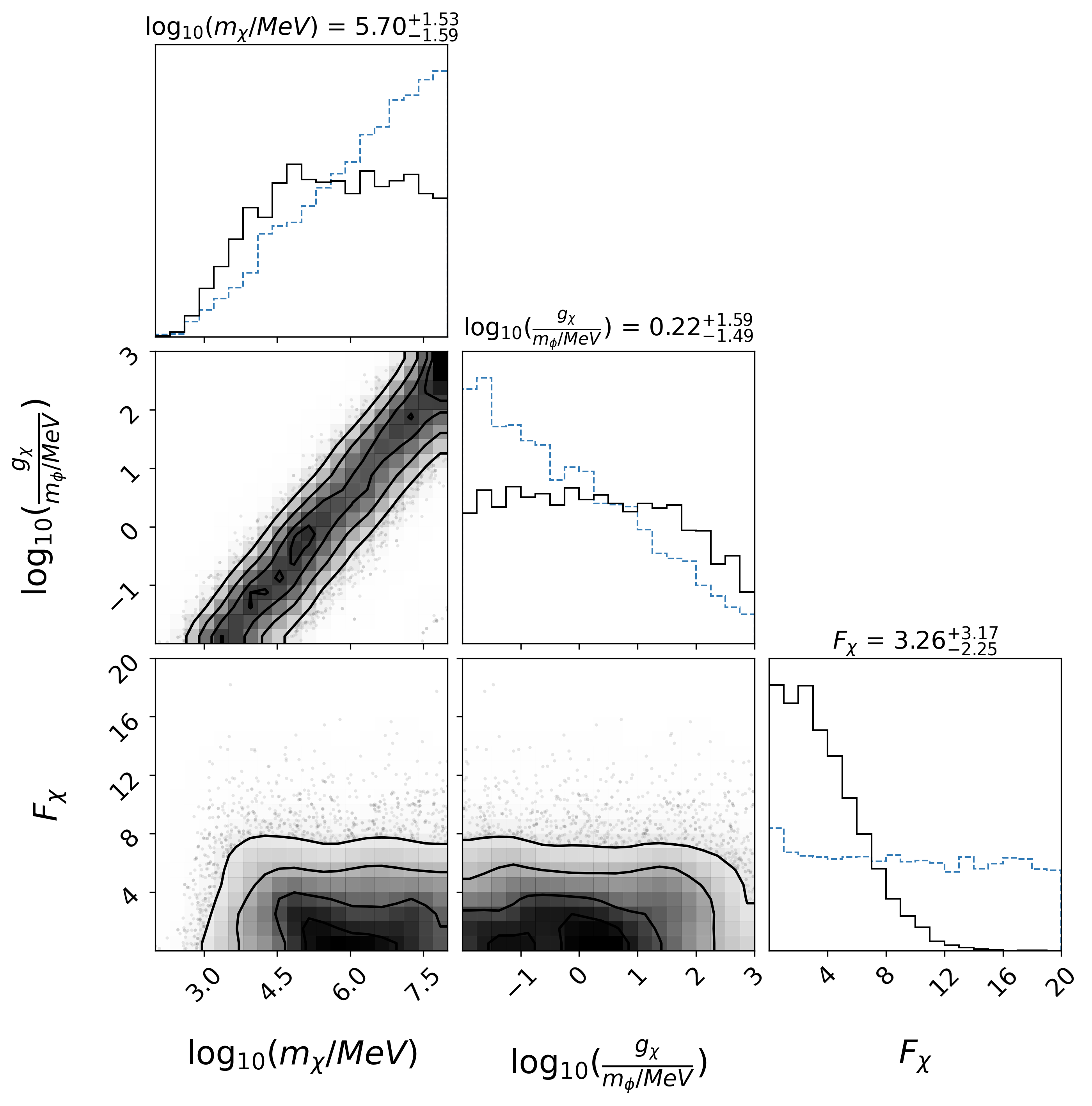}
  \label{fig5:sub3}
\end{subfigure}%
\begin{subfigure}{.5\textwidth}
  \centering
  \includegraphics[width=\textwidth]{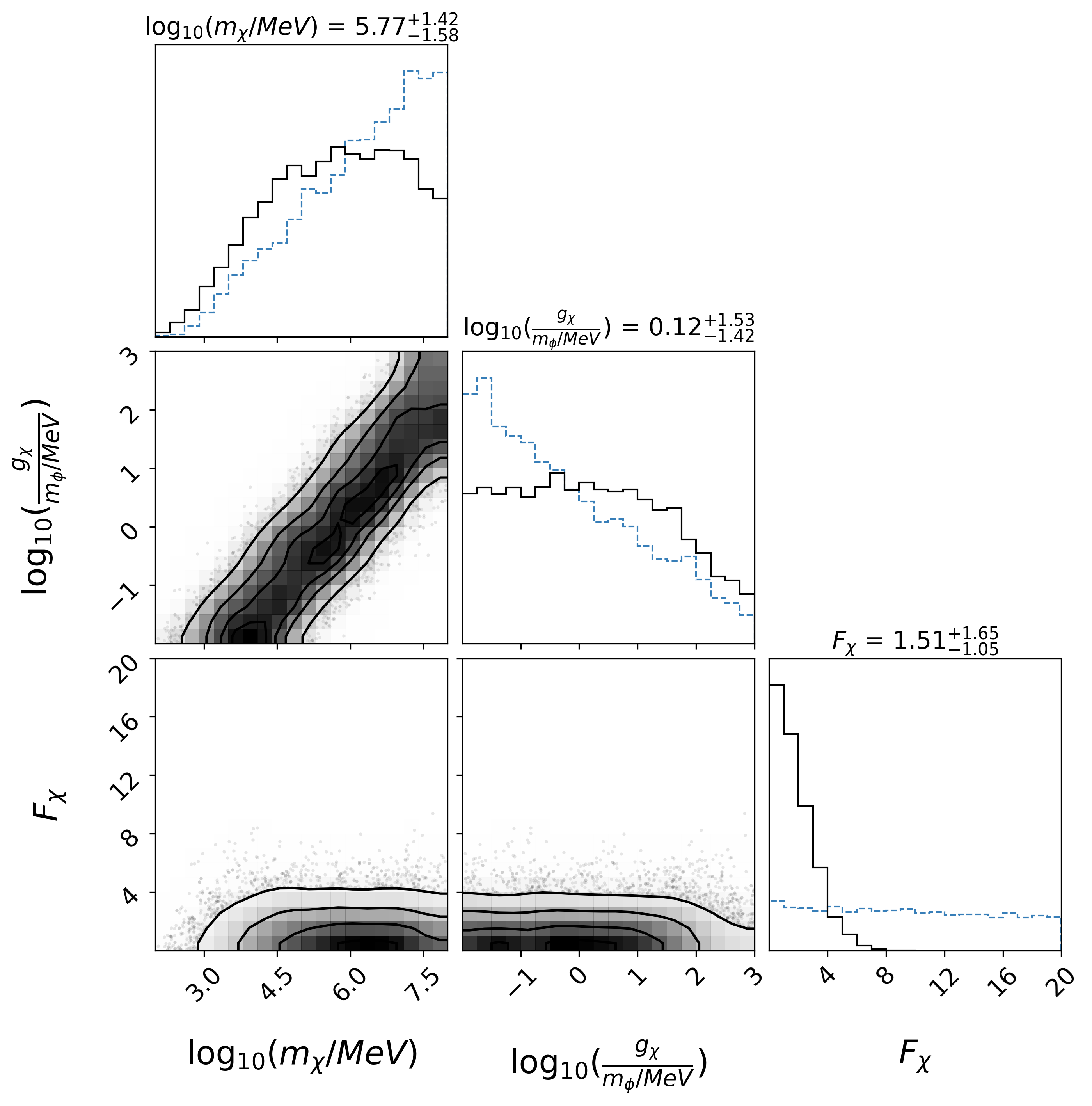}
  \label{fig5:sub4}
\end{subfigure}
\caption{(Left two panels) \textit{Future} scenario with fixed baryonic matter EoS posterior corner plot; (Right two panels) \textit{Future-X} scenario with fixed baryonic matter EoS posterior corner plot. Where the priors, posteriors, diagonal figure titles, and ground truth models for each respective panels follows the same flow as in Fig.~\ref{fig3}. The $F_\chi$ 1-D posteriors of all quadrants are tighter than the posteriors of $F_\chi$ in Fig.~\ref{fig3} and have significant peaks around their respective ground truth models. Also, the stripe shown in the $\log_{10}\boldsymbol{(}g_\chi/(m_\phi/\mathrm{MeV})\boldsymbol{)}$ vs. $\log_{10}(m_\chi/\mathrm{MeV})$ 2-D posteriors is noticeably wider for the ``No ADM" ground truth models than the ``ADM core" models.}
\label{fig5}
\end{figure*}
\subsection{\textit{Future}/\textit{Future-X}: Fixed baryonic EoS}\label{Fixed BM EoS}
We now consider a fixed baryonic matter EoS. Performing a Bayesian analysis on the dark matter parameter space with a fixed baryonic matter EoS is similar to previous analyses \cite{Apran2020,Miao_2022}. It is hard to envisage a situation where we would be able to constrain the baryonic matter EoS using neutron stars observationally independent of the role of any ADM contribution, since ADM will also affect gravitational wave observations. 
However, there are prospects for reducing the uncertainty on the baryonic EoS from lab/ heavy ion collision measurements, which may be less affected by ADM \citep[see, e.g.,][]{CREX2022,PPREX2021,Huth2022,Keller2022}. Here we also explore what \textit{Future-X} might be able to deliver in comparison to a \textit{Future} scenario. For the prior distribution of the bosonic ADM parameters, we use the priors shown in Fig.~\ref{fig2} because the same ADM parameters were sampled over as in the varying baryonic matter case.

Figure~\ref{fig5} shows the posterior distributions of the ADM parameters. The 2-D posteriors show that the ratio of $\log_{10}\boldsymbol{(}g_\chi/(m_\phi/\mathrm{MeV})\boldsymbol{)}$ and $\log_{10}(m_\chi/\mathrm{MeV})$ is constrained to a diagonal stripe. Here we notice, the high $m_\chi$ and low $g_\chi/m_\phi$ regime -- which we characterize by the region in which $m_\chi \gtrsim 10^6 $ MeV and $g_\chi/m_\phi \lesssim 10^{-1} \mathrm{MeV^{-1}}$ -- is effectively disfavored for both \textit{Future} and \textit{Future-X} scenarios. In order to accumulate any appreciable ADM mass in this regime, the ADM central energy density needs to be much larger than the baryonic matter central energy density. For example, to have $F_\chi = 0.05\%$ we need $ \epsilon_\chi(r=0) \sim 10^8 \epsilon_B(r=0)$ for a baryonic matter EoS with a maximum mass of $\approx 2.3 \, \Msun$. The resulting mass and radius of an ADM admixed neutron star from this example has a similar radius to the baryonic matter neutron star, but the mass is significantly smaller. In this example, the maximum mass of the neutron star produced from baryonic matter EoS is reduced to $0.374 \, \Msun$. Therefore, this regime is disfavored because it does not satisfy the imposed constraint that the produced neutron stars must have a mass of at least $1\, \Msun$. The $1 \Msun$ constraint is motivated by the theoretical description of a neutron star early in its evolution \citep[see, e.g.,][]{Strobel_1999}. 

The 2-D posteriors of the ratio between the effective self-repulsion strength and the bosonic ADM particle mass also show that the stripe widens for the ``No ADM" model in both \textit{Future} and \textit{Future-X} scenarios. The diagonal stripe widens in the ``No ADM" model because the ground truth mass-fraction is 0\%. This allows for pairs of $g_\chi/m_\phi$ and $m_\chi$, which can produce neutron stars above the 1 $\Msun$ constraint for sufficiently low mass-fractions, to be given a significant likelihood evaluation. However, these same pairs of points would be given a low likelihood in for the ``ADM core" model because they do not produce neutron stars above the 1 $\Msun$ constraint for mass-fractions near the ground truth value of 7\%.

According to the 1-D posterior histograms of $m_\chi$ and $g_\chi/m_\phi$, we find that both parameters do not have a significant peak. From this observation, we conclude that $m_\chi$ and $g_\chi/m_\phi$ are individually unconstrained quantities in the remaining parameter space. However, Fig.~\ref{fig5} shows that the mass-fraction 1-D posteriors do have a significant peak and contrasts significantly to the varying baryonic EoS case. The $F_\chi$ 1-D posterior is more Gaussian-like in shape than its 1-D posterior in Fig.~\ref{fig3} because fixing the baryonic EoS reduces the number of parameters sampled over from 10 to 3. Therefore, any degeneracies between the baryonic EoS parameters and the ADM mass-fraction are eliminated, which results in the mass-fraction 1-D posterior to be more Gaussian-like. The Gaussian-like shape of the 1-D $F_\chi$ posteriors of both \textit{Future} and \textit{Future-X} shows that if the baryonic EoS is fixed, i.e., better understood than it is now, tight constraints on $F_\chi$ can be imposed.

For \textit{Future-X}, Fig.~\ref{fig5} shows both of the 1-D $F_\chi$ posteriors are centered around the $F_\chi$ ground truth value. Figure~\ref{fig5} also shows that the 1-D $F_\chi$ posterior distributions of the ``ADM core'' and ``No ADM'' models have $4\sigma$ deviations of $\approx \pm 9.8\%$ and $\approx\pm 7.6\%$, respectively. The \textit{Future} scenario also displays that both 1-D $F_\chi$ posteriors are centered around the ground truth value, but ``ADM core'' and ``No ADM'' models have $4\sigma$ deviations of $\approx \pm 19.3\%$ and $\approx \pm 14.7\%$, respectively. These results show that \textit{Future-X} can provide tighter 1-D $F_\chi$ posteriors than \textit{Future}, highlighting \textit{Future-X}'s ability to place better constraints on the ADM mass-fraction than \textit{Future}.

\section{Conclusion}\label{Conclusion}
In this work, we adapted the Bayesian analysis of \cite{Raaijmakers2019} to include a bosonic ADM core component in neutron stars. We have considered the bosonic ADM model of \cite{Nelson_2018}, which describes bosonic ADM particles with self-repulsion. We performed Bayesian inferences of the bosonic ADM particle mass $m_\chi$, effective self-repulsion strength $g_\chi/m_\phi$, and mass-fraction $F_\chi$, which uses synthetic mass-radius posteriors for two different instrumentation scenarios, namely the \textit{Future} (``NICER'') and \textit{Future-X} (``STROBE-X'') scenarios. Interestingly, the synthetic sources used for \textit{Future-X} suggest that if two neutron stars with the same mass but different radii were measured at the 2\% uncertainty level, one could conclude that one neutron star has ADM while the other does not, since both neutron stars share the same baryonic matter EoS. A scenario like this would hint at variation in the amount of ADM in neutron stars. 

In the cases where the baryonic matter EoS is varied, we find the 2-D posteriors of $\log_{10}\boldsymbol{(}g_\chi/(m_\phi/\mathrm{MeV})\boldsymbol{)}$ and $\log_{10}(m_\chi/\mathrm{MeV})$ are identical in all plots. For the mass-fraction, we find the 1-D posteriors of the ``ADM core" model are not centered around the ground truth mass-fraction value in both scenarios. We also find that the 1-D $F_\chi$ posteriors are nearly identical in all four quadrants. These results show that, within the current uncertainties of the baryonic EoS, \textit{Future} and \textit{Future-X} will not be able to determine if ADM present in neutron stars nor will they be able to provide constraints on the bosonic ADM parameter space. Although, if the actual ADM mass-fraction is not very small, the ratio of $g_\chi/m_\phi$ and $m_\chi$ is well constrained, which is demonstrated by the wider PDF contours in the low $F_\chi$ regime of Fig.~\ref{fig3a}. 

In the energy density-pressure plane, we find the statistical uncertainty in the inference of the neutron star EoS is widened when the possibility of an ADM core is taken into account. For example, at $\mathrm{log_{10}}(\epsilon \, \mathrm{cm^3/g}) = 14.71$, the \textit{Future} baryonic EoS uncertainty increased by 12.19\% and 19.62\% for both ground truth models when including ADM, while \textit{Future-X} showed a 17.79\% and 24.98\% increase. This shows that the current uncertainties of the baryonic EoS are being underestimated because the possibility of dark matter in neutron stars is being ignored. We recommend that projections of future datasets -- and data analysis pipelines -- should account for the possibility that neutron stars contain ADM cores.

The ADM posteriors differ significantly from the posteriors of the varying baryonic EoS case when the baryonic matter EoS is fixed. Both \textit{Future} and \textit{Future-X} show that the high ADM particle mass ($m_\chi \gtrsim 10^6$ MeV) and low effective self-interaction strength ($g_\chi/m_\phi \lesssim 0.1 \, \mathrm{MeV^{-1}}$) regime is disfavored due to the observational and theoretical constraint that neutron stars must have a mass above $1\, \Msun$. The diagonal stripe in the 2-D posteriors of $m_\chi$ and $g_\chi/m_\phi$ is wider for the ``No ADM" model in both scenarios because the ground truth mass-fraction for the ``No ADM" model is lower than the ``ADM core" model ground truth mass-fraction. This allows for points in the $g_\chi/m_\phi$-$m_\chi$ plane that do not satisfy the $1\, \Msun$ constraint for mass-fractions close to the ``ADM core" ground truth to be given a higher likelihood in the ``No ADM" model. Although the 2-D posteriors show that the ratio of $m_\chi$ and $g_\chi/m_\phi$ is constrained, we find the individual quantities are not well constrained. However, the 1-D $F_\chi$ posteriors are well constrained due to their Gaussian-like shapes and are centered around the ground truth value. We find the 1-D $F_\chi$ posteriors of \textit{Future} have $4\sigma$ deviations of $\approx \pm 19.3\%$ and $\approx \pm 14.7\%$ for the ``ADM core'' and ``No ADM'' ground truth models, respectively. For the \textit{Future-X} 1-D $F_\chi$ posteriors, we find the ``ADM core'' model has a $4\sigma$ deviation of $\approx \pm 9.8\%$ and the ``No ADM' model has a $4\sigma$ deviation of $\approx \pm 7.6\%$. This highlights that \textit{Future-X} will be able to provide stronger constraints on the bosonic ADM mass-fraction than \textit{Future}. Our results show that, if the baryonic EoS is known better than it is currently, $F_\chi$ can be tightly constrained and constraints on the ratio of the bosonic ADM particle mass and effective self-interaction strength can be inferred. This work demonstrates NASA missions, NICER and STROBE-X, will not be able to provide constraints on bosonic ADM under the current uncertainties of the baryonic EoS. However, if the baryonic EoS uncertainties are tightened independent of ADM, NICER and STROBE-X demonstrate an ability to constrain the ratio of $m_\chi$ and $g_\chi/m_\phi$ and $F_\chi$, but not the individual quantities of $m_\chi$ and $g_\chi/m_\phi$ in the \cite{Nelson_2018} bosonic ADM model.   

These results are physically reasonable because the bosonic ADM particle mass and effective self-interaction significantly impact the distribution of the ADM core. This is most evident in the 1-D posteriors of Fig.~\ref{fig5} since it does not show significant peaks. Therefore, neither individual parameter has a significant effect on the mass and radius of the admixed neutron star, which our Bayesian analysis is done in reference to. Within the uncertainties considered, $m_\chi$ and $g_\chi/m_\phi$ have indistinguishable effects on the mass-radius and they primarily effect the distribution of ADM in the neutron star. Although individually $m_\chi$ and $g_\chi/m_\phi$ are unconstrained, their ratio is constrained because it determines the density of the ADM particles in the ADM radius, which affects the central energy density needed to reach a given $F_\chi$. This is demonstrated by the solid white triangulated region in 2-D posteriors of $\log_{10}\boldsymbol{(}g_\chi/(m_\phi/\mathrm{MeV})\boldsymbol{)} \, vs. \, \log_{10}(m_\chi/\mathrm{MeV})$ in Fig.~\ref{fig5}. The low mass and high effective self-interaction strength regime has a low posterior density evaluation because it produces ADM halos for any ADM central energy density. This is different from the high mass and low effective self-interaction strength regime, which has a low posterior evaluation because it allows for many heavy ADM particles to be contained within a small ADM core radius, which causes the ADM central energy density needed to achieve a fixed $F_\chi$ to be $\mathcal{O}(10^{22} \, g/cm^3)$. ADM central energy densities of this magnitude produce neutron stars below the $1 \, \Msun$ constraint. Lastly, it is physically consistent that the bosonic ADM mass-fraction is the most constrained parameter in this analysis because it is directly related to the total mass of ADM, which affects the mass and radius of the admixed neutron star.  

We have also shown the value of performing full inference runs on the ADM parameter space, rather than drawing conclusions only from the effects that each ADM parameter has on the mass-radius relation. Degeneracies between the parameters can confound the effects of the ADM parameters, which would make it difficult to draw meaningful conclusions on any of the parameters if an inference was not done. Our inference method shows a degeneracy, but also displays an ability to provide constraints on ADM parameters which may otherwise not be found. Finally, we would like to point out, as we did in Sec.~\ref{Varying BM EoS}, that this work is a best case scenario for future STROBE-X observations and end of mission NICER observations. Therefore, future STROBE-X and NICER mass-radius measurements will likely have larger uncertainties than we used in this work, and our constraints will relax accordingly.

Future work will account for ADM halos by appropriately modifying the ray-tracing models used by the NICER collaboration and how PPM is conducted. As \citet{Miao_2022} has shown, this can help us fully understand the effects ADM halos have on neutron mass-radius measurements and the potential halos have in providing new insights on the constraints of the ADM parameter space. Future work will also include a scatter of the data sources around the ground truth model. Although we did not include scatter in our sources due the added arbitrariness, \citet{Greif19} shows that including scatter around the ground truth models can help us study real mass-radius data in the most general way. The combination of full inferences with scatter and ADM halos will allow for general and accurate statements about the constraints on the ADM parameter space. More broadly, our work adds to the growing literature which shows how to use these inference methods to constrain dark matter models that interact with neutron star interiors.
\begin{center}
    \textbf{ACKNOWLEDGMENTS}
    \end{center}

We acknowledge Ann Nelson for her foundational contributions to this work and helpful conversations with C.P.W. about the purpose of model-building. We thank Djuna Croon, Rebecca Leane, and Nathan Musoke for helpful discussions. We also thank Sharon Morsink, Slavko Bogdanov, Paul Ray, and Tom Maccarone for offering feedback on the manuscript. Lastly, we thank the anonymous referee for comments that helped improve this work. A.L.W. acknowledges support from ERC Consolidator grant No. 865768 AEONS. C.P.W. acknowledges all the administrative and facilities staff at the University of New Hampshire. The contributions of C.P.W. and N.R. were supported by NASA grant No. 80NSSC22K0092. G.R. is grateful for financial support from the Nederlandse Organisatie voor Wetenschappelijk Onderzoek (NWO) through the Projectruimte and VIDI grants (Nissanke).

\appendix
\section{\label{appendix A}Derivation of the bosonic ADM equation of state }
The Lagrangian of the model described in Sec.~\ref{ADMBackground} which  yields the EoS in Eq.~(\ref{ADM_EoS1}) and Eq.~(\ref{ADM_EoS2}) can be written down, in units of $\hbar = c = 1$, as
\begin{multline}
        \mathcal{L}_\chi = -\sqrt{-g} \Big( D_\mu^* \chistar D^\mu \chi  + m_\chi^2 \chistar\chi +\frac{1}{2} m_\phi^2 \phi_\mu \phi^\mu\\
    + \frac{1}{4}Z_{\mu\nu} Z^{\mu \nu} - g_B \phi_\mu J_B^\mu \Big),
\end{multline}
Before we begin the derivation, we make a few approximations to for ease of calculations. As discussed in Sec.~\ref{ADMBackground}, we approximate $g_B$ to be $g_B \ll g_\chi$. Therefore, we may neglect the $g_B$ in the Lagrangian. Next, we approximate the spacetime to be flat since the effects of gravity are negligible on the inverse length scales of neutron stars. To convince ourselves that this a good approximation, we follow the reasoning in \cite{Sagun_2022_Boson}. 

Consider the gradient of the $g_{tt}$ component of the metric. Since we are working in a spherically symmetric spacetime, then one can write the gradient as 
\begin{equation} \label{dgttdr}
   \frac{d g_{tt}}{d r} =  \frac{-2 g_{tt}}{P + \epsilon} \frac{d P}{dr}, 
\end{equation}
where $P$ is the pressure and $\epsilon$ is the energy density. Consider the stellar surface, then $g_{tt} = -1 + \frac{2G M}{Rc^2}$ and p = 0, where $M$ is the total gravitational mass of the star and $R$ is its radius. The $g_{tt}$ component then allows Eq.~\ref{dgttdr} to be written as
\begin{equation}
    \frac{d g_{tt}}{d r} = -\frac{2 G M}{c^2R^2}.
\end{equation}
In the limit where $g_{tt} = 0$, which implies $\frac{2GM}{Rc^2} = 1$, we now have that 
\begin{equation}
   \abs{\frac{d g_{tt}}{d r}} < \frac{1}{R}.
\end{equation}
Therefore, for a star of radius $10 \, \mathrm{km}$, a typical radius of a neutron star, and we consider a spherical layer of thickness $\delta r = 10^{-3} \, \mathrm{km}$, which is large enough to treat dark matter thermodynamically while also ensuring that we have
\begin{equation}
    \abs{\frac{d g_{tt}}{d r} }\delta r < \frac{\delta r}{R} = 10^{-4}. 
\end{equation}
Therefore the derivatives of the metric are small compared to the spatial scales of a neutron star, so we can approximate our spacetime as being flat. The final form of the Lagrangian is given as
\begin{multline}
    \mathcal{L}_\chi = -\sqrt{-g}\big(D_\mu^* \chistar D^\mu \chi  - m_\chi^2 \chistar\chi \\
    -\frac{1}{2} m_\phi^2 \phi_\mu \phi^\mu-  \frac{1}{4}Z_{\mu\nu} Z^{\mu \nu} \big).
\end{multline}
By variations of the action, we obtain the equations of motions Eqs \ref{3}, \ref{4}, and \ref{5} in Sec.~\ref{ADMBackground}.
In the mean-field approximation, we assume separability and make the stationary scalar field ansatz of 
\begin{equation}
    \chi(r,t) = (Ae^{ikr}+Be^{-ikr})e^{-i\w t},
\end{equation}
where $A$ and $B$ are real constants, $k$ is the wave-number, and $\w$ is the eigen frequency of the spherically symmetric bound state of the scalar field. Since we are treating the vector field as a classical field, then $\phimu = (\phinaught,0)$, where $\phinaught$ is a constant. Plugging in the expressions for the fields $\chi$ and $\phimu$ into the equations of motion and letting $k=0$, we find 
\begin{equation}
    \w =  g_\chi \phinaught \pm m_\chi.
\end{equation}
Since $m_\chi$ could be larger than $g_\chi \phi_0$ and we demand that $\omega>0$ then, we will take the $``+"$ solution. Since $\phimu = (\phi_0,0)$ the equation of motion of $\phimu$ gives us a relation between $\phi_0$ and $\chistar \chi$ to be
\begin{equation}
    \mphi^2 \phinaught = 2g_\chi m_\chi \chistar \chi.
\end{equation}
Next we calculate the conserved current, $J^\mu$, since $J_0 = n_\chi$ where $n_\chi$ is the ADM number density. 
\begin{align}
    J^\mu & = \frac{i}{\sqrt{-g}}\left( \chistar \frac{\partial \mathcal{L}_\chi}{\partial \nabla_\mu \chistar} - \frac{\partial \mathcal{L}_\chi}{\partial \nabla_\mu \chi} \chi  \right) \\ 
    &= -i \big(\chistar \nabla^\mu \chi -  (\nabla^\mu \chistar)\chi\big) + 2 g_\chi \phi^\mu \chistar \chi.
\end{align}

Writing out the $J_0 = g_{0\nu}J^\nu$ element using equation for $\w$ yields that
\begin{align}
n_\chi &= 2\w \chistar \chi - 2 g_\chi \phinaught \chistar \chi  = 2 \mchi \chistar \chi\\
&\implies \phinaught = \frac{g_\chi}{\mphi^2} n_\chi.
\end{align}
To write down the stress-energy tensor one needs to vary the action with respect to the metric \cite{Carroll2004}. In the mean-field approximation, we find the stress-energy tensor to be 
\begin{multline}
    T_{\mu \nu} = 2 D_\mu^* \chistar D_\nu \chi - g_{\mu\nu} D_\rho^* \chistar D^\rho \chi  \\+ m_\phi^2 \big( \phi_\mu \phi_\nu - \frac{1}{2} g_{\mu\nu}\phi_\rho \phi^\rho\big) - g_{\mu\nu}m_\chi^2 \chistar \chi
\end{multline}
The ADM energy density, $\epsilon_\chi$, is determined from $T_{00}$ component, and the ADM pressure, $p_\chi$, is determined by $T_{ii}$. Writing out the respective components of the stress-energy tensor, the ADM EoS, with $\hbar$ and $c$ restored, is 
\begin{align}
    \epsilon_\chi &= \mchi c^2 n_\chi +\frac{1}{2} \frac{g_\chi^2}{\mphi^2} \frac{\hbar^3}{c} n_\chi^2\\
    p_\chi &= \frac{1}{2} \frac{g_\chi^2}{\mphi^2} \frac{\hbar^3}{c} n_\chi^2.
\end{align}
This is Eq.~(\ref{ADM_EoS1}) and Eq.~(\ref{ADM_EoS2}).

\bibliography{bibliography}

\end{document}